\documentclass[10pt,twocolumn]{IEEEtran}
\usepackage{mathrsfs}
\usepackage{color}
\usepackage{epsfig}
\usepackage{amsmath}
\usepackage{cite}
\usepackage{amsfonts, amssymb}
\usepackage{multirow}
\usepackage{array}
\usepackage{diagbox}
\newtheorem{theorem}{Theorem}

\newtheorem{lemma}{Lemma}

\newtheorem{assumption}{Assumption}
\newtheorem{remark}{Remark}

\def\BibTeX{{\rm B\kern-.05em{\sc i\kern-.025em b}\kern-.08em
T\kern-.1667em\lower.7ex\hbox{E}\kern-.125emX}}

\allowdisplaybreaks

\begin{document}
\title{Minimum-Variance Recursive State Estimation for 2-D Systems: When Asynchronous
 Multi-Channel Delays meet Energy Harvesting Constraints
\thanks{This work was supported by the National Natural Science Foundation of China (No. 62250056), the Natural Science Foundation of Shandong Province (No.ZR2021MF069,ZR2021JQ24).}
}
\author{~Yu Chen and ~Wei Wang$^{*}$
\thanks{Y.~Chen and W.~Wang  are with the School of Control Science and Engineering, Shandong University, Jinan 250061, P.~R.~China (e-mail: ychen\_2000@163.com; w.wang@sdu.edu.cn).}
}
\maketitle

\begin{abstract}
    This paper is concerned with the state estimation problem for two-dimensional systems with asynchronous multi-channel delays and energy harvesting constraints. In the system, each smart sensor has a certain probability of harvesting energy from the external environment, the authorized transmission between the sensor and the remote filter is contingent upon the current energy level of the sensor, which results in intermittent transmission of observation information. Addressing the issue of incomplete observation information due to asynchronous multi-channel delays, a novel approach for observation partition reconstruction is proposed to convert the delayed activated observation sequences into equivalent delay-free activated observation sequences. Through generating spatial equivalency validation, it is found that the reconstructed delay-free activated observation sequences contain the same information as the original delayed activated observation sequences. Based on the reconstructed activated observation sequence and activated probability, a novel unbiased $\hbar$+1-step recursive estimator is constructed. Then, the evolution of the probability distribution of the energy level is discussed. The estimation gains are obtained by minimizing the filtering error covariance. Subsequently, through parameter assumptions, a uniform lower bound and a recursive upper bound for the filtering error covariance are presented. And the monotonicity analysis of activated probability on estimation performance is given. Finally, the effectiveness of the proposed estimation scheme is verified through a numerical simulation example.
\end{abstract}

\begin{IEEEkeywords}
    Two-dimensional system, asynchronous multi-channel delay, energy harvesting constraint, observation partition reconstruction , $\hbar$+1-step recursive estimator.
\end{IEEEkeywords}

\section{Introduction}\label{sec:1}
    With the development of modern industrial manufacturing, the requirements for multivariate analysis are getting higher and higher. Two-dimensional (2-D) systems have attracted widespread attention because they propagate along two independent directions. Bidirectional propagation, as a distinctive feature of 2-D systems, distinguishes it from classic one-dimensional (1-D) systems that evolve in a single direction. In \cite{Yaz_AML01}, the formulation of a 2-D system Roesser model have been given through the description of a multi-dimensional iterative circuit. Subsequently, a more general FM model of the 2-D systems have been given  in \cite{Petersen12}, and discussed its structural properties such as reachability, observability and internal stability accordingly. Since then, various 2-D systems have been studied and applied to batch processing systems, environmental monitoring systems, and iterative learning tracking systems, etc \cite{Givone_ITC72,Fornasini_MST78,Song_SMC23}. So far, a lot of work has been done on issues such as system stabilization, filtering and fault estimation, and controller design in the 2-D framework \cite{Wang_CAA22,Li_SMC23,Chinimilli_ISJ19,Liu_TAC,Xing_JRNC21,Zhu_ATO22,Xie_ISP02}.
	
    The filtering problem has been one of the fundamental problems in the field of control and signal processing, with the aim of restoring the system state based on observations corrupted by various undesirable network-induced phenomena (e.g. delays, quantization, biases, degradations, censorings, and outliers, etc). It is worth noting that in actual engineering applications, noise is ubiquitous and enters process devices or observation outputs in different forms. Once these noises are not adequately handled, they may seriously affect the estimation performance and even cause the estimation algorithms to fail. Thereby, various filtering algorithms have been developed to deal with different types of noise in 2-D systems, depending on the type of noise (including but not limited to random noise, unknown-but-bounded noises). To date, a vast number of research results have been reported in the literature (see e.g. \cite{Yang_SMC21,Lu_AJC22,Yang_TAC19,Wang_RNC24,Jacobson_JMAA74}).
	
    Regarding the filtering problem in the network environment, due to data conflicts, transmission errors and network congestion, delays may occur in multi-channel transmission of measurement information in a limited-bandwidth communication network. Several techniques commonly used to solve delays (e.g., system augmentation,  linear matrix inequality) have been applied to the 2-D systems state estimation problem. For example, in \cite{Liu_TAC23}, the randomly sensor delay filtering problem using stochastic Kronecker delta functions to represent actual measurements has been studied. By using system augmentation techniques to transform the estimation problem with randomly sensor delays into an equivalent delay-free estimation problem. In \cite{Wang_TNSE22}, the 2-D complex network state estimation with randomly varying sensor delays has been studied. In \cite{Zhu_TAC23}, the robust $H_{\infty}$ filtering with state-varying delays has been studied. By using linear matrix inequality techniques, a sufficient condition for $H_{\infty}$ performance analysis have been developed. In \cite{Nayyar_TAC13}, the distributed $H_{\infty}$ estimation problem of 2-D stochastic delay systems has been studied. By using stochastic analysis and linear matrix inequality techniques, sufficient conditions for the estimation error to be globally asymptotically stable on the mean square have been established. It is intuitive and effective to use system augmentation technology to deal with the estimation problem of delays systems. However, the system augmentation increases system dimensions and creates unnecessary computational overhead. And the results obtained by linear matrix inequality techniques are obviously conservative and relatively fuzzy theoretical analysis. Therefore, developing an estimation algorithm that can handle delay issues without incurring extra computational overhead is the main motivation of our research.
	
    Due to the rapid development and widespread application of wireless sensor networks, communication energy consumption needs to be considered during signal transmission. Traditionally, each sensor device could be powered by a rechargeable/replaceable outdoor mobile power source or a stationary power source, but this may technically impractical given economic and practical environmental factors. To cope with the limited energy supply of the sensor, guarantee authorized transmission between the sensor and the remote estimator, in \cite{Leung_TSP15,Huang_TAC17}, by utilizing energy harvesting technology, the sensor is replenished with energy from the external environment through various sources (e.g. solar, wind, etc.). It is worth noting that the energy obtained by the sensor from the external environment is random, which means that the authorized transmission between the sensor and the remote filter is also random, that is, the communication between the sensor and the remote filter is intermittent. Therefore, it is a seemingly interesting but natural idea to explore the probability distribution of energy level and the statistical properties of transmission probability. So far, the problem of state estimation with energy harvesting constraints has received a lot of attention in 1-D systems and some satisfactory results have been obtained \cite{Zhao_JRNC17,Liang_JRNC14,Wang_ATO23,Tarighati_TSP17,Chen_TCB23}. But the corresponding results in 2-D systems have not been fully studied, let alone the case where asynchronous multi-channel delays and stochastic nonlinearities are considered simultaneously, which is the main motivation of our study.
	
    The four fundamental challenges we will encounter are summarized below: 1) How to develop an efficient state estimation scheme in the presence of asynchronous multi-channel delays, energy harvesting constraints, and stochastic nonlinearities? 2) When the energy storage state is known, how to calculate the probability distribution of energy level under the evolution law? 3) How to analyze the boundedness of the filtering error covariance? 4) How to measure the impact of activation probability on estimation performance? Therefore, the main aim of this paper is to present satisfactory solutions to these identified challenges. {\it The main contributions of this paper can be highlighted as follows: 1) A novel $\hbar$+1-step estimator framework is established for 2-D systems with asynchronous multi-channel delays, energy harvesting constraints, and stochastic nonlinearities. 2) Through probabilistic analysis methods, an explicit expression of the energy level probability distribution is derived. 3) Through mathematical induction and parameter assumptions, the recursive upper bound of the filtering error covariance is presented. 4) The monotonicity analysis of the activated probability on the estimation performance is given.}

    The remainder of this work is arranged as follows. In Section \ref{sec:2}, the considered 2-D system estimation problem with multi-channel delays is formulated. In Section \ref{sec:3}, based on the partition reconstruction activation observation, an $\hbar$+1-step unbiased recursive filter is proposed, and the estimation gain is appropriately designed to minimize the filtering error covariance. In Section \ref{sec:4}, an analysis of the estimation performance is given. Conclusions are lastly drawn in Section \ref{sec:6}.

    {\it Notations:} The notation used here is normative. $\mathcal{E}\{\cdot\}$ represents the mathematical expectation. The scrip `$\mathrm{T}$' represents the transpose of vectors or matrices. $\mathbf{R}^n$ stands the $n$-dimensional Euclidean space. $\delta(i,j)$ is the Kronecker delta function with $\delta(i,j)$ being unity for $i=j$ but zero elsewhere. $col_{0\leq c \leq s}\{e_c\}$ represents a column vector with $e_0,e_1,\cdots,e_s$ as an element. $diag_{0\leq c \leq s}\{e_c\}$ stands for a diagonal matrix with $e_0,e_1,\cdots,e_s$ as the diagonal elements. $[0,s]$ represents the set $\{0,1,\cdots,s\}$. $\mathbf{1}_s$ represents the $s$-dimensional row vector with all elements being scalar 1. $\text{Prob}\{\cdot\}$ means the occurrence probability of the event `$\cdot$'.
\section{Problem formulation}\label{sec:2}
    \subsection{System setup}
        Consider the following 2-D system over a finite horizon $i,j\in[0,\varkappa]$ with $\varkappa>0$ being a given scalar:
		\begin{equation}\label{CW2a:1}
			\left\{
				\begin{aligned}
					x(i,j)&=A_1(i,j-1)x(i,j-1)+A_2(i-1,j)\\
					&\quad\times x(i-1,j)+g(x(i,j-1),\xi(i,j-1))\\
                    &\quad+g(x(i-1,j),\xi(i-1,j))+B_1(i,j-1)\\
                    &\quad\times w(i,j-1)+B_2(i-1,j)w(i-1,j)\\
					z_{(s)}(i,j)&=C_{(s)}(i,j)x(i,j)+h_{(s)}(x(i,j),\gamma(i,j))\\
                    &\quad+v_{(s)}(i,j)
				\end{aligned}
			\right.
		\end{equation}
        where $x(i,j)\in\mathbf{R}^n$ and $z_{(s)}(i,j)\in\mathbf{R}^{p_s}(s\in[0,\hbar])$ are the system state and the $s$+1-th measured output, respectively; $g(x(i,j),\xi(i,j))$ and $h_{(s)}(x(i,j),\gamma(i,j))$ are stochastic nonlinear function; $w(i,j)\in\mathbf{R}^m$ and $v_{(s)}(i,j)\in\mathbf{R}^{p_s}$ are zero-mean Gaussian white noise with covariance $Q(i,j)\geq0$ and $R_{(s)}(i,j)>0$. For $c\in[1,2],s\in[0,\hbar]$, $A_c(i,j), B_c(i,j),$ and $C_{(s)}(i,j)$ are known shift-varying parameters with applicable dimensions.
		
		The system state $x(i,j)$ is observed by $\hbar$+1 different unreliable channels with delay as described by
		\begin{align}\label{CW2a:2}
			y_{(s)}(i,j)=&z_{(s)}(i-\imath_s,j-\jmath_s)
		\end{align}
        Without loss of generality, the asynchronous multi-channel delay $\imath_s$ and $\jmath_s$ being a strictly increasing order: $0=\imath_0=\jmath_0<\imath_1\leq\jmath_1<\cdots<\imath_{\hbar}\leq\jmath_{\hbar}$.

		For the convenience of narration, the following three assumptions have been made in this paper.
		\begin{assumption}
             The initial conditions for the 2-D systems \eqref{CW2a:1} and \eqref{CW2a:2} are given below:
			\begin{align*}
				&\mathcal{E}\big\{x(i,0)\big\}=\hat{x}_u(i,0),\quad \mathcal{E}\big\{x(0,j)\big\}=\hat{x}_u(0,j) \\
				&Cov\big\{x(i,0),x(t,0)\big\}=\Xi(i,0)\delta(i,t)\\
				&Cov\big\{x(0,j),x(0,t)\big\}=\Xi(0,j)\delta(j,t)\\
				&Cov\big\{x(i,0),x(0,j)\big\}=\Xi(0,0)\delta(i,0)\delta(0,j)
			\end{align*}
            where $\hat{x}_u(i,0), \hat{x}_u(0,j), \Xi(i,0),$ and $\Xi(0,j)$ are known shift-varying parameters with applicable dimensions.
		\end{assumption}
		\begin{assumption}
            The stochastic nonlinear functions $g(x(i,j),$ $\xi(i,j))$ and $h(x(i,j),\gamma(i,j))$ possess the following statistical properties:
			\begin{small}
            \begin{equation}\label{CW2a:3}
				\left\{
					\begin{aligned}
						&g(0,\xi(i,j))=0,\ h_{(s)}(0,\gamma(i,j))=0\\
						&\mathcal{E}\Bigg\{\begin{bmatrix}
							g(x(l,k),\xi(l,k)) \\
							h_{(s)}(x(l,k),\gamma(l,k))
						\end{bmatrix}\Big|x(i,j)
						\Bigg\}=0\\
						&\qquad (l,k)\in \big\{(l_1,k_1)|l_1>i\ \text{or}\ k_1>j\big\}\cup (i,j)\\
						&\mathcal{E}\Bigg\{\begin{bmatrix}
								g(x(i,j),\xi(i,j)) \\
								h_{(s)}(x(i,j),\gamma(i,j))
							\end{bmatrix}\begin{bmatrix}
								g(x(l,k),\xi(l,k)) \\
								h_{(s)}(x(l,k),\gamma(l,k))
							\end{bmatrix}^\mathrm{T}\Big|x(i,j)
							\Bigg\}\\
						&\qquad =\sum_{\mu=1}^{r}\Delta_{\mu}x^\mathrm{T}(i,j)\Sigma_{\mu}x(i,j)\delta(i,l)\delta(j,k)
					\end{aligned}
				\right.
			\end{equation}
            \end{small}
            where $\xi(i,j)$ and $\gamma(i,j)$ are zero-mean random sequences with covariance $\epsilon^2I$ and $\iota^2I$, $r>0$ is a given integer, $\Delta_{\mu}=diag\big\{\Delta_{g,\mu},\Delta_{h_{(s),\mu}}\big\}$, $\Sigma_{\mu}=diag\big\{\Sigma_{g,\mu},\Sigma_{h_{(s)},\mu}\big\}$ with $\mu\in[1,r]$, $\Delta_{g,\mu}, \Delta_{h_{(s),\mu}}, \Sigma_{g,\mu}$, and $\Sigma_{h_{(s)},\mu}$ are known parameters.
		\end{assumption}
        \begin{assumption}
            The random variables $w(i,j), v_{(s)}(i,j),\\ \xi(i,j)$, and $\gamma(i,j)$ are white and mutually uncorrelated, which are also independent of the initial state $x(i,0)$ and $x(0,j) $ with $i,j\in[0,\varkappa], s\in[0,\hbar]$.
        \end{assumption}
        \begin{remark}
            The stochastic nonlinearities characterized by \eqref{CW2a:3} have been first proposed by Jacobson in the seminal work \cite{Fornasini_MST78}, and then the state estimation problem of stochastic nonlinearity for 2-D systems has been considered in \cite{Liang_IS16}. Nonlinear phenomena are often encountered in practical systems and are considered as one of the important sources that complicate system analysis and synthesis. Randomly occurring nonlinear phenomena, often referred to as stochastic nonlinearities, are very general and can include many special cases, such as state-dependent multiplicative noise and random variables whose power depends on the sign of the system state.
		\end{remark}
		\begin{remark}
            In scenarios such as communication lines, network transmission systems, electronic signal processing systems, and urban traffic management systems, information transmission is a complex and critical process. These systems often rely on communication networks with limited bandwidth to handle and transmit large volumes of data. However, due to bandwidth limitations, delays commonly occur during the transmission process, a phenomenon generally referred to as communication delay. This can negatively affect the performance and efficiency of the systems.
		\end{remark}
    \subsection{Energy harvesting constraints}
		\begin{figure}[!ht]
			\centering
			\includegraphics [scale=0.6]{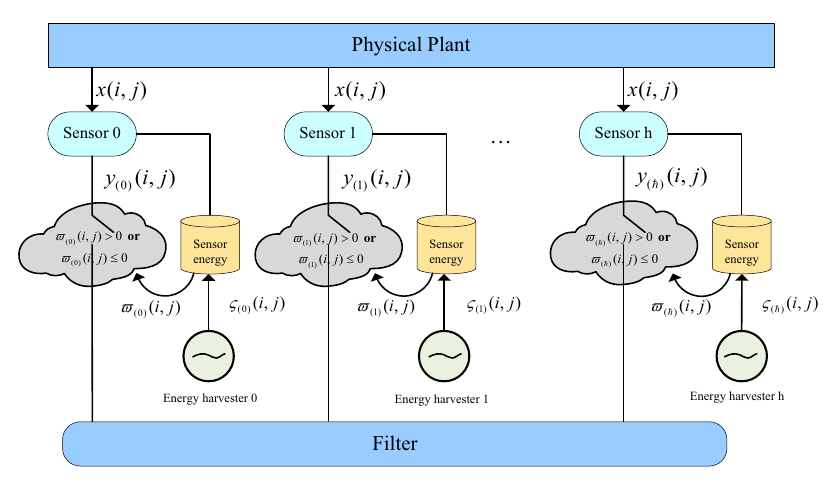}
			\caption{Target system configuration with sensor harvesting constraints.}
			\label{fig.1}
		\end{figure}
        Configuration of target system with sensor harvesting constraints is shown in Fig.~\ref{fig.1}, where the sensor with energy harvesting harvests external energy from the environment, and data transmission (between the sensor and the remote filter) can only be authorized when the current energy storage is sufficient.
		
        The energy storage state of sensor $s$ at the horizon $(i,j)$ is represented by  $\varpi_{(s)}(i,j)\in\{0,1,2,\cdots,M_{(s)}\}$, where $M_{(s)}$ is the maximum of energy stored. The amount of harvested energy $\varsigma_{(s)}(i,j)$ is an independent and identically distributed (i.i.d) stochastic process with the following probability distribution
		\begin{align}\label{CW2b:1}
			\text{Prob}(\varsigma_{(s)}(i,j)=\flat)=p_{(s)}(\flat),\quad \flat=0,1,2,\cdots,
		\end{align}
		where $0\leq p_{(s)}(\flat)\leq 1$ and $\sum_{\flat=0}^{\infty}p_{(s)}(\flat)=1$.
		
        When sensor $s$ energy storage state $\varpi_{(s)}(i,j)>0$, the sensor consumes 1 unit of energy to transmit the measured information to the remote filter. Therefore, the energy storage dynamics of sensor $s$ can be described as
		\begin{equation}\label{CW2b:2}
			\left\{
				\begin{aligned}
					&\varpi_{(s)}(i+1,j+1)=min\big\{\varpi_{(s)}(i+1,j)+\varpi_{(s)}(i,j+1)\\
					&\qquad+\varsigma_{(s)}(i+1,j)+\varsigma_{(s)}(i,j+1)-\mathbf{1}_{\{\varpi_{(s)}(i+1,j)>0\}}\\
					&\qquad-\mathbf{1}_{\{\varpi_{(s)}(i,j+1)>0\}}, M_{(s)}\big\}\\
					&\varpi_{(s)}(i,0)=m_{(s),1}(i), \varpi_{(s)}(0,j)=m_{(s),2}(j)
				\end{aligned}
			\right.
		\end{equation}
        where $i,j\in[0,\varkappa], m_{(s),1}(i),m_{(s),2}(j)\in\mathbb{N}$. The activated observation received by the remote filter can be represented as
		\begin{align}\label{CW2b:3}
			\bar{y}_{(s)}(i,j)=\mathbf{1}_{\{\varpi_{(s)}(i,j)>0\}}y_{(s)}(i,j)
		\end{align}
		where $\mathbf{1}_{\{\varpi_{(s)}(i,j)>0\}}$ is an indicator function satisfying
		\begin{equation*}
			\mathbf{1}_{\{\varpi_{(s)}(i,j)>0\}}=\left\{
				\begin{aligned}
					&1,\quad \varpi_{(s)}(i,j)>0\\
					&0,\quad \varpi_{(s)}(i,j)\leq0
				\end{aligned}
			\right.
		\end{equation*}
		\begin{remark}
            It is worth noting that energy harvesting models are divided into two categories. One type is deterministic models, and the other type is stochastic models (including time-correlated models and time-uncorrelated models). According to different models, energy harvesting constraint can be modeled by Poisson process and Bernoulli process, respectively. The relevant descriptions in 1-D can be seen in \cite{Wang_SMC20,Hu_SMC21,Shen_ATO19}. In this paper, we assume that the probability distribution of sensor energy collection is known, and model the energy storage dynamics of sensors under the bidirectional evolution characteristics of a 2-D system for the first time.
		\end{remark}
    \subsection{Original activated observation description }
        Let $\mathscr{Y}(l,k)$ denotes the original activated observation of the state subject to energy harvesting constraints at the horizon $(l,k)$. For the sake of clarity, define the following sets
		\begin{align*}
			&\mathcal{M}_0^+=\{(l,k)|\imath_0\leq l< \imath_1,\jmath_1\leq k\leq j\},\nonumber \\
			&\mathcal{M}_0^\circ\ =\{(l,k)|\imath_0\leq l< \imath_1,\jmath_0\leq k< \jmath_1\},\nonumber \\
			&\mathcal{M}_0^-\ =\{(l,k)|\imath_1\leq l\leq i,\jmath_0\leq k< \jmath_1\},\nonumber\\
			&\qquad\qquad\qquad\qquad\qquad \vdots \nonumber\\
            &\mathcal{M}_{\hbar-1}^+=\{(l,k)|\imath_{\hbar-1}\leq l< \imath_{\hbar},\jmath_{\hbar}\leq k\leq j\},\nonumber \\
            &\mathcal{M}_{\hbar-1}^\circ=\{(l,k)|\imath_{\hbar-1}\leq l< \imath_{\hbar},\jmath_{\hbar-1}\leq k< \jmath_{\hbar}\},\nonumber \\
            &\mathcal{M}_{\hbar-1}^-=\{(l,k)|\imath_{\hbar}\leq l\leq i,\jmath_{\hbar-1}\leq k< \jmath_{\hbar}\},\nonumber\\
			&\mathcal{M}_{\hbar}^\circ\quad =\{(l,k)|\imath_{\hbar}\leq l\leq i,\jmath_{\hbar}\leq k\leq j\},\nonumber\\
            &\mathcal{M}_s=\mathcal{M}_s^+\cup \mathcal{M}_s^\circ\cup \mathcal{M}_s^-(s\in[0,\hbar-1]), \mathcal{M}_\hbar=\mathcal{M}_{\hbar}^\circ.\nonumber
		\end{align*}
	
		The activated observation can be describe as
			\begin{align}\label{CW2c:1}
			\mathscr{Y}(l,k)=
			\begin{bmatrix}
				\bar{y}_{(0)}(l,k)\\
				\bar{y}_{(1)}(l,k)\\
				\vdots\\
				\bar{y}_{(s)}(l,k)
			\end{bmatrix}
		\end{align}
        where $(l,k)\in \mathcal{M}_s$ for $s\in[0,\hbar]$.
		\begin{remark}
            Due to the effects of multi-channel observation delays and energy harvesting constraints, it can be seen from \eqref{CW2c:1} that the activated observation sequence at each horizons are different, and observation within the same horizon reflect different state information. Thereby we cannot develop a filter of shape \cite{Oubaidi_IJDC23} by using activated observation subject to energy harvesting constraints, which goes against the idea of using as many observation as possible to estimate the state. In order to make full use of the activated observation that reflect the same state information and develop a recursive form of filter, the activated observation sequence will be reconstructed below.
		\end{remark}
\section{The main results}\label{sec:3}
	\subsection{Activated observation reconstruction}
        So as to detailed investigate the estimation problem with multi-channel observation delays and energy harvesting constraints, we only consider the case of $(i,j)\in\big\{(i_0,j_0)|\imath_\hbar\leq i_0,\jmath_\hbar\leq j_0\big\}$, other cases can be handled in the same way. For brevity, denote $i_s=i-\imath_{\hbar-s},j_s=j-\jmath_{\hbar-s}$ and define the following sets
		\begin{align*}
			&\mathcal{N}_0^\circ=\{(l,k)|0\leq l\leq i_0,0\leq k\leq j_0\},\nonumber\\
			&\mathcal{N}_1^-=\{(l,k)|0\leq l\leq i_0,j_0<k\leq j_1\},\nonumber\\
			&\mathcal{N}_1^\circ=\{(l,k)|i_0<l\leq i_1,j_0< k\leq j_1\},\nonumber\\
			&\mathcal{N}_1^+=\{(l,k)|i_0<l\leq i_1,0\leq k\leq j_0\},\nonumber\\
			& \qquad\qquad\qquad\quad \vdots\nonumber\\
			&\mathcal{N}_{\hbar}^-=\{(l,k)|0\leq l\leq i_{\hbar-1},j_{\hbar-1}<k\leq j_{\hbar}\},\nonumber\\
			&\mathcal{N}_{\hbar}^\circ=\{(l,k)|i_{\hbar-1}<l\leq i_{\hbar},j_{\hbar-1}< k\leq j_{\hbar}\},\nonumber\\
			&\mathcal{N}_{\hbar}^+=\{(l,k)|i_{\hbar-1}<l\leq i_{\hbar},0\leq k\leq j_{\hbar-1}\},\nonumber\\
            &\mathcal{N}_0=\mathcal{N}_0^\circ,\mathcal{N}_s=\mathcal{N}_s^-\cup \mathcal{N}_s^\circ \cup \mathcal{N}_s^+(s\in[1,\hbar]).\nonumber
		\end{align*}
		
		The activated observation can be reconstructed as follows:
		\begin{align}\label{CW3a:1}
			y_s(l,k)=
			\begin{bmatrix}
				\bar{y}_{(0)}(l,k)\\
				\bar{y}_{(1)}(l+\imath_1,k+\jmath_1)\\
				\vdots\\
				\bar{y}_{(s)}(l+\imath_s,k+\jmath_s)
			\end{bmatrix}
		\end{align}
		where $(l,k)\in \mathcal{N}_{\hbar-s}$ for $s\in[0,\hbar]$.	
	
		It is easy to know that $y_s(l,k)$ satisfy
		\begin{align}\label{CW3a:2}
			y_s(l,k)=&\mathbf{1}_{\{\varpi_s(l,k)>0\}}\Big(C_s(l,k)x(l,k)+v_s(l,k)\nonumber\\
			&+h_s(x(l,k),\gamma(l,k))\Big)
		\end{align}
        where $\mathbf{1}_{\{\varpi_s(l,k)>0\}}=diag_{0\leq c\leq s}\big\{\mathbf{1}_{\{\varpi_{(c)}(l+\imath_c,k+\jmath_c)>0\}}\big\},\\ C_s(l,k)=col_{0\leq c \leq s}\big\{C_{(c)}(l,k)\big\},v_s(l,k)=col_{0\leq c \leq s}\big\{v_{(c)}(l,\\ k)\big\}, h_s(x(l,k),\gamma(l,k))=col_{0\leq c \leq s}\big\{h_{(c)}\big(x(l,k),\gamma(l,k)\big)\big\}$, $v_s(l,k)$ is zero-mean white noise with covariance matrices $R_s(l,k)=
        diag_{0\leq c\leq s}\big\{R_{(c)}(l,k)\big\}$. Obviously, there is delay-free existed in \eqref{CW3a:2}. The following lemma will show that the reconstructed activated observation sequence and the original activated observation sequence have the same information.
		\begin{lemma}\label{lem1}
            The linear space spanned by the reconstructed activation observation sequences and the linear space spanned by the original activated observation sequences are equivalent, i.e.
			\begin{equation*}
				\begin{aligned}
                    &\mathscr{L}\Big\{\big\{y_{\hbar}(l,k)|(l,k)\in \mathcal{N}_0\};\{y_{\hbar-1}(l,k)|(l,k)\in \mathcal{N}_1\};\cdots;\\
                    &\{y_0(l,k)|(l,k)\in \mathcal{N}_{\hbar}\big\}\Big\}=\mathscr{L}\big\{\mathscr{Y}(l,k)|(l,k)\in \mathcal{M}_0\cup \mathcal{M}_1\cup\\
					&\cdots \cup \mathcal{M}_{\hbar}\big\}.
				\end{aligned}
			\end{equation*}
		\end{lemma}
        \begin{IEEEproof}
            On the ground of \eqref{CW2b:2}-\eqref{CW2c:1}, it is easy to show that for $(l,k)\in M_{\hbar}$, $\mathcal{Y}(l,k)$ is a linear combination of $y_s(l,k)$,
            \begin{small}
			\begin{equation*}
				\begin{aligned}
					\mathscr{Y}(l,k)&=\begin{bmatrix} 
						I_{p_0}\\ \mathbf{0}_{(p_1+\cdots+p_{\hbar})\times p_0}
					\end{bmatrix}y_0(l,k)\\
					&+\begin{bmatrix}
                    \mathbf{0}_{p_0\times p_0} & \mathbf{0}_{p_1\times p_1} \\ \mathbf{0}_{p_1\times p_0} & I_{p_1} \\ \mathbf{0}_{(p_2+\cdots+p_{\hbar})\times p_0} & \mathbf{0}_{(p_2+\cdots+p_{\hbar})\times p_1}    
					\end{bmatrix}y_1(l-\imath_1,k-\jmath_1)
					\\&+\cdots\\
					&+
					\begin{bmatrix}
                        \mathbf{0}_{(p_0+\cdots+p_{\hbar-1})\times (p_0+\cdots+p_{\hbar-1})} & \mathbf{0}_{(p_0+\cdots+p_{\hbar-1})\times p_{\hbar}}  \\ \mathbf{0}_{p_{\hbar}\times (p_0+\cdots+p_{\hbar-1})} & I_{p_{\hbar}}
					\end{bmatrix}\\
				&\quad\times y_{\hbar}(l-\imath_{\hbar},k-\jmath_{\hbar}).
				\end{aligned}
			\end{equation*}
            \end{small}
			
			Moreover, $y_s(l,k)$ is also a linear combination of $\mathcal{Y}(l,k),$
            \begin{small}
			\begin{align*}
				&y_0(l,k)=\begin{bmatrix}
					I_{p_0} & \mathbf{0}_{p_0\times (p_1+\cdots+p_{\hbar})}
				\end{bmatrix}\mathscr{Y}(l,k),\nonumber\\
				&y_1(l,k)=\begin{bmatrix}
					I_{p_0} & \mathbf{0}_{p_0 \times (p_1+\cdots+p_{\hbar})} \\
					\mathbf{0}_{p_1\times p_0} & \mathbf{0}_{p_1\times (p_1+\cdots+p_{\hbar})}
				\end{bmatrix}\mathscr{Y}(l,k)\nonumber\\
				&\qquad\quad+\begin{bmatrix}
                    \mathbf{0}_{p_0\times p_0} & \mathbf{0}_{p_0\times p_1} & \mathbf{0}_{p_0\times (p_2+\cdots+p_{\hbar})} \\ \mathbf{0}_{p_1\times p_0} & I_{p_1} & \mathbf{0}_{p_1\times (p_2+\cdots+p_{\hbar})}
				\end{bmatrix}\\
				&\qquad\qquad\times\mathscr{Y}(l+\imath_1,k+\jmath_1),\nonumber\\
				&\cdots\cdots\nonumber \\
				&y_{\hbar}(l,k)=\begin{bmatrix}
					I_{p_0} & \mathbf{0}_{p_0\times (p_1+\cdots+p_{\hbar})} \\
                    \mathbf{0}_{(p_1+\cdots+p_{\hbar})\times p_0} & \mathbf{0}_{(p_1+\cdots+p_{\hbar})\times (p_1+\cdots+p_{\hbar})}
				\end{bmatrix}\\
				&\qquad\qquad\times\mathscr{Y}(l,k)\nonumber\\
				&+\begin{bmatrix}
                    \mathbf{0}_{p_0\times p_0} & \mathbf{0}_{p_0\times p_1} &  \mathbf{0}_{p_0\times (p_2+\cdots+p_{\hbar})} \\
					\mathbf{0}_{p_1\times p_0} & I_{p_1} & \mathbf{0}_{p_1\times (p_2+\cdots+p_{\hbar})} \\
                    \mathbf{0}_{(p_2+\cdots+p_{\hbar})\times p_0} & \mathbf{0}_{(p_2+\cdots+p_{\hbar})\times p_1} & \mathbf{0}_{(p_2+\cdots+p_{\hbar})\times (p_2+\cdots+p_{\hbar})}
				\end{bmatrix}\\
				&\qquad\qquad\times\mathscr{Y}(l+\imath_1,k+\jmath_1)\nonumber\\
				&+\cdots+\begin{bmatrix}
                    \mathbf{0}_{(p_0+\cdots+p_{\hbar-1})\times(p_0+\cdots+p_{\hbar-1})} & \mathbf{0}_{(p_0+\cdots+p_{\hbar-1})\times p_{\hbar}} \\
					\mathbf{0}_{p_{\hbar}\times(p_0+\cdots+p_{\hbar-1})} & I_{p_{\hbar}}
				\end{bmatrix}\\
				&\qquad\qquad\times\mathscr{Y}(l+\imath_{\hbar},k+\jmath_{\hbar}).
			\end{align*}
            \end{small}
            It can be proved the case of $(l,k)\in \mathcal{M}_0\cup \mathcal{M}_1\cup \cdots \cup \mathcal{M}_{\hbar-1}$ in the same way. This completes the proof.
		\end{IEEEproof}
		\begin{remark}
            The new observation sequence, $\{y_s(l,k)|(l,k)\\ \in \mathcal{M}_0\cup\mathcal{M}_1\cup \cdots \cup \mathcal{M}_{\hbar}\} $ is named as reconstructed activated observation sequence of $\{\mathscr{Y}(l,k)|(l,k)\in \mathcal{N}_0\cup \mathcal{N}_1\cup \cdots \cup \mathcal{N}_{\hbar}\}$. It is obvious from Lemma~\ref{lem1} that the reconstructed activated observation sequence contains the same information as the original activated observation sequence.
		\end{remark}
    \subsection{Filter development}
        According to the area division of the reconstructed activated observation, an $\hbar$+1-step recursive filter is proposed as follows:
		
		\begin{subequations}\label{CW3b:1}
			$\mathbf{Step\  1}$: For $(l,k)\in \mathcal{N}_0$
			\begin{align}
				\hat{x}_p^1(l,k)=&A_1(l,k-1)\hat{x}_u^1(l,k-1)\nonumber\\
				&+A_2(l-1,k)\hat{x}_u^1(l-1,k)\label{CW3b:1a}\\
				\hat{x}_u^1(l,k)=&\hat{x}_p^1(l,k)+K_1(l,k)\nonumber\\
				&\times\big[y_{\hbar}(l,k)-\varrho_{\hbar}(l,k)C_{\hbar}(l,k)\hat{x}_p^1(l,k)\big]\label{CW3b:1b}
			\end{align}
		\end{subequations}
        where $\hat{x}_p^1(l,k)$ is the one-step prediction of the state $x(l, k)$, $\hat{x}_u^1(l,k)$ is the corresponding updated estimate with the initial conditions $\hat{x}_u^1(l,0)=\hat{x}_u(l,0)$ and $\hat{x}_u^1(0,k)=\hat{x}_u(0,k)$.
		
		\begin{subequations}\label{CW3b:2}
			$\mathbf{Step\  2}$: For $(l,k)\in \mathcal{N}_1$
			\begin{align}
				\hat{x}_p^2(l,k)=&A_1(l,k-1)\hat{x}_u^2(l,k-1)\nonumber\\
				&+A_2(l-1,k)\hat{x}_u^2(l-1,k)\label{CW3b:2a}\\
				\hat{x}_u^2(l,k)=&\hat{x}_p^2(l,k)+K_2(l,k)\nonumber\\
                &\times\big[y_{\hbar-1}(l,k)-\varrho_{\hbar-1}(l,k)C_{\hbar-1}(l,k)\hat{x}_p^2(l,k)\big]\label{CW3b:2b}
			\end{align}
		\end{subequations}
        where $	\hat{x}_p^2(l,k)$ is the one-step prediction of the state $x(l,k)$, $\hat{x}_u^2(l,k)$ is the corresponding updated estimate with the initial conditions $\hat{x}_u^2(l,0)=\hat{x}_u(l,0)(i_0+1\leq l \leq i_1),\hat{x}_u^2(l,j_0)=\hat{x}_u^1(l,j_0)(0 \leq l \leq i_0)$ and $\hat{x}_u^2(0,k)=\hat{x}_u(0,k)(j_0+1 \leq k \leq j_1), \hat{x}_u^2(i_0,k)=\hat{x}_u^1(i_0,k)(0 \leq k \leq j_0)$.
		
		$\qquad\qquad\qquad\qquad\qquad \vdots$
		
		\begin{subequations}\label{CW3b:3}
			$\mathbf{Step\  \hbar+1}$: For $(l,k)\in \mathcal{N}_{\hbar}$
			\begin{align}
				\hat{x}_p^{\hbar+1}(l,k)=&A_1(l,k-1)\hat{x}_u^{\hbar+1}(l,k-1)\nonumber\\
				&+A_2(l-1,k)\hat{x}_u^{\hbar+1}(l-1,k)\label{CW3b:3a}\\
				\hat{x}_u^{\hbar+1}(l,k)=&\hat{x}_p^{\hbar+1}(l,k)+K_{\hbar+1}(l,k)\nonumber\\
				&\times\big[y_{0}(l,k)-\varrho_{0}(l,k)C_{0}(l,k)\hat{x}_p^{\hbar+1}(l,k)\big]\label{CW3b:3b}
			\end{align}
		\end{subequations}
        where $	\hat{x}_p^{\hbar+1}(l,k)$ is the one-step prediction of the state $x(l,k)$, $\hat{x}_u^{\hbar+1}(l,k)$ is the corresponding updated estimate with the initial conditions $\hat{x}_u^{\hbar+1}(l,0)=\hat{x}_u(l,0)(i_{\hbar-1}+1\leq l \leq i_{\hbar}),\\ \hat{x}_u^{\hbar+1}(l,j_{\hbar-1})=\hat{x}_u^{\hbar}(l,j_{\hbar-1})(0 \leq l \leq i_{\hbar-1})$ and $\hat{x}_u^{\hbar+1}(0,k)\\=\hat{x}_u(0,k) (j_{\hbar-1}+1 \leq k \leq j_{\hbar}), \hat{x}_u^{\hbar+1}(i_{\hbar-1},k)=\hat{x}_u^{\hbar}(i_{\hbar-1},k)\\(0 \leq k \leq j_{\hbar-1})$.
		
        In \eqref{CW3b:1}-\eqref{CW3b:2}, $K_1(l,k), K_2(l,k),\cdots, K_{\hbar+1}(l,k)$ are estimation gains to be determined, $\varrho_{s}(l,k)$ represent the activated probability and
		\begin{align*}
            \varrho_{s}(l,k)=diag_{0\leq c \leq s}\big\{\mathcal{E}[\mathbf{1}_{\{\varpi_{(c)}(l+\imath_c,k+\jmath_c)>0\}}]\big\}
		\end{align*}
		\begin{remark}
            The innovation information $y_s(l,k)-\varrho_s(l,k)\\ C_s(l,k)\hat{x}_p^{\hbar+1-s}(l,k)$ for $s\in[0,\hbar]$ are utilized to update the estimate. The estimator of the above form \eqref{CW3b:1}-\eqref{CW3b:2} are proposed based on the reconstructed activated observation sequence because of its similarity to the estimator structure specified in \cite{Zou_CSSP04}. So $\hbar$+1-step recursive filter is also called Kalman-type filter based on reconstructed activated observation.
		\end{remark}
	
		For the $\hbar$+1-step recursive filter, denote the prediction errors and filtering errors,
		\begin{align}
			&\tilde{x}_p^{\gamma}(l,k)\triangleq x(l,k)-\hat{x}_p^{\gamma}(l,k)\label{CW3b:4}\\
			&\tilde{x}_u^{\gamma}(l,k)\triangleq x(l,k)-\hat{x}_u^{\gamma}(l,k)\label{CW3b:5}
		\end{align}
		Then, it follow from \eqref{CW2a:1}-\eqref{CW2a:2} and \eqref{CW3b:1}-\eqref{CW3b:3} that
		
		\begin{subequations}\label{CW3b:6}
			$\mathbf{Step\  1}$: For $(l,k)\in \mathcal{N}_{0}$
            \begin{small}
			\begin{align}
				\tilde{x}_p^1(l,k)=&A_1(l,k-1)\tilde{x}_u^1(l,k-1)+A_2(l-1,k)\nonumber\\
				&\times\tilde{x}_u^1(l-1,k)+g(x(l,k-1),\xi(l,k-1))\nonumber\\
				&+g(x(l-1,k),\xi(l-1,k))+B_1(l,k-1)\nonumber\\
				&\times w(l,k-1)+B_2(l-1,k)w(l-1,k)\label{CW3b:6a}\\
				\tilde{x}_u^1(l,k)=&[I-K_1(l,k)\varrho_{\hbar}(l,k)C_{\hbar}(l,k)]\tilde{x}_p^1(l,k)\nonumber\\
				&-K_1(l,k)\Big\{\big[\mathbf{1}_{\varpi_{\hbar}(l,k)>0}-\varrho_{\hbar}(l,k)\big]\nonumber\\
				&\times C_{\hbar}(l,k)x(l,k)+\big[\mathbf{1}_{\varpi_{\hbar}(l,k)>0}\big]\nonumber\\
				&\times \big(v_{\hbar}(l,k)+h_{\hbar}(x(l,k),\gamma(l,k))\big)\Big\}\label{CW3b:6b}
			\end{align}
            \end{small}
		\end{subequations}
	
		\begin{subequations}\label{CW3b:7}
			$\mathbf{Step\  2}$: For $(l,k)\in \mathcal{N}_{1}$
            \begin{small}
			\begin{align}
				\tilde{x}_p^2(l,k)=&A_1(l,k-1)\tilde{x}_u^2(l,k-1)+A_2(l-1,k)\nonumber\\
				&\times\tilde{x}_u^2(l-1,k)+g(x(l,k-1),\xi(l,k-1))\nonumber\\
				&+g(x(l-1,k),\xi(l-1,k))+B_1(l,k-1)\nonumber\\
				&\times w(l,k-1)+B_2(l-1,k)w(l-1,k)\label{CW3b:7a}\\
				\tilde{x}_u^2(l,k)=&[I-K_1(l,k)\varrho_{\hbar-1}(l,k)C_{\hbar-1}(l,k)]\tilde{x}_p^2(l,k)\nonumber\\
				&-K_2(l,k)\Big\{\big[\mathbf{1}_{\varpi_{\hbar-1}(l,k)>0}-\varrho_{\hbar-1}(l,k)\big]\nonumber\\
				&\times C_{\hbar-1}(l,k)x(l,k)+\big[\mathbf{1}_{\varpi_{\hbar-1}(l,k)>0}\big]\nonumber\\
				&\times \big(v_{\hbar-1}(l,k)+h_{\hbar-1}(x(l,k),\gamma(l,k))\big)\Big\}\label{CW3b:7b}
			\end{align}
            \end{small}
		\end{subequations}
	
		$\qquad\qquad\qquad\qquad\qquad \vdots$
		
		\begin{subequations}\label{CW3b:8}
			$\mathbf{Step\  \hbar+1}$: For $(l,k)\in \mathcal{N}_{\hbar}$
            \begin{small}
			\begin{align}
				\tilde{x}_p^{\hbar+1}(l,k)=&A_1(l,k-1)\tilde{x}_u^{\hbar+1}(l,k-1)+A_2(l-1,k)\nonumber\\
				&\times\tilde{x}_u^{\hbar+1}(l-1,k)+g(x(l,k-1),\xi(l,k-1))\nonumber\\
				&+g(x(l-1,k),\xi(l-1,k))+B_1(l,k-1)\nonumber\\
                &\times w(l,k-1)+B_2(l-1,k)w(l-1,k)\label{CW3b:8a}\\
                \tilde{x}_u^{\hbar+1}(l,k)=&[I-K_{\hbar+1}(l,k)\varrho_{0}(l,k)C_{0}(l,k)]\tilde{x}_p^{\hbar+1}(l,k)\nonumber\\
				&-K_{\hbar+1}(l,k)\Big\{\big[\mathbf{1}_{\varpi_{0}(l,k)>0}-\varrho_{0}(l,k)\big]\nonumber\\
				&\times C_{0}(l,k)x(l,k)+\big[\mathbf{1}_{\varpi_{0}(l,k)>0}\big]\nonumber\\
				&\times \big(v_{0}(l,k)+h_{0}(x(l,k),\gamma(l,k))\big)\Big\}\label{CW3b:8b}
			\end{align}
            \end{small}
		\end{subequations}
	
        This paper investigated 2-D systems estimation with asynchronous multi-channel delays and energy harvesting constraints, and proposed an $\hbar$+1-step recursive estimator in form of \eqref{CW3b:1}-\eqref{CW3b:3}, striving to achieve the following three conditions:
		
        (1) Calculation the activated probability to ensure that the reconstructed activated observation can calculate innovation to update estimate;
		
        (2) Construct an upper bound in the sense of spectral norm to constrain the filtering error covariance, in addition, the trace of filtering error covariance is minimizing  by designing estimation gains appropriately;
		
		(3) Analyze the impact of probability distribution of energy levels on the estimation performance index.

    \subsection{Calculation of the activated probability}
        In this subsection, the approximate calculation of the activated probability is given through the upper bound of the energy level probability distribution.
		\begin{lemma}
            For the sensor energy level $\varpi_{(s)}(l,k)$ given by \eqref{CW2b:2}, denote $\digamma_{(s),\flat}(l,k) \triangleq\text{Prob}\{\varpi_{(s)}(l,k)=\flat\}$. Denote the probability distribution of the energy level by
			\begin{align*}
                \digamma_{(s)}(l,k)\triangleq\big[\text{Prob}\{\varpi_{(s)}(l,k)=0\} \cdots \text{Prob}\{\varpi_{(s)}(l,k)=M_{(s)}\}\big]^\mathrm{T}
			\end{align*}
			with the initial conditions
            \begin{small}
			\begin{equation*}
				\left\{
				\begin{aligned}
                    &\digamma_{(s)}(l,0)=\big[\ \underbrace{0 \cdots 0}_{m_{(s),1}(l)}\quad 1\ \underbrace{0 \cdots 0}_{M_{(s)}-m_{(s),1}(l)}\ \big]^\mathrm{T}\\
                    &\digamma_{(s)}(0,k)=\big[\ \underbrace{0 \cdots 0}_{m_{(s),2}(k)}\quad 1\ \underbrace{0 \cdots 0}_{M_{(s)}-m_{(s),2}(k)}\ \big]^\mathrm{T}
				\end{aligned}
				\right.
			\end{equation*}
            \end{small}
			The probability distribution $\digamma_{(s)}(l,k)$ satisfy the following inequality
			\begin{align}\label{lem2:1}
				\digamma_{(s)}(l,k)\leq\mathcal{F}_{(s)}(l,k)
			\end{align}
            Let $\mathcal{F}_{(s)}(l,k)\triangleq\big[\text{Prob}\{\breve{\varpi}_{(s)}(l,k)=0\} \cdots \text{Prob}\{\breve{\varpi}_{(s)}(l,k)=M_{(s)}\}\big]^\mathrm{T}$ and $\mathcal{F}_{(s),\flat}(l,k) \triangleq\text{Prob}\{\breve{\varpi}_{(s)}(l,k)=\flat\}$. The recursive evolution of $\mathcal{F}_{(s)}(l,k)$ as follows:
			\begin{equation}\label{lem2:2}
				\left\{
				\begin{aligned}
                    &\mathcal{F}_{(s)}(l+1,k+1)= Q_{(s)}\big[\mathcal{F}_{(s)}(l+1,k)+\mathcal{F}_{(s)}(l,k+1)\big]\\
					&\mathcal{F}_{(s)}(l,0)=\digamma_{(s)}(l,0), \mathcal{F}_{(s)}(0,k)=\digamma_{(s)}(0,k)				
				\end{aligned}
				\right.
			\end{equation}
            where $Q_{(s)}=[\Theta_{(s),1}^\mathrm{T},\Theta_{(s),2}^\mathrm{T}]^\mathrm{T}$, $\Theta_{(s),1}$ and $\Theta_{(s),2}$ are given as shown in \eqref{le2:5} and \eqref{le2:6}.
			\begin{figure*}[ht!]
				\centering
				\begin{small}
                \begin{align}
					&\Theta_{(s),1}=\begin{bmatrix}
						\Theta_{(s),1}(0,0) & \Theta_{(s),1}(0,1) & 0 & \cdots & 0 \\
						\Theta_{(s),1}(1,0) & \Theta_{(s),1}(1,1) & \Theta_{(s),1}(1,2) & \cdots & 0 \\
						\vdots & \vdots & \vdots & \ddots & \vdots \\
                        \Theta_{(s),1}(M_{(s)}-1,0) & \Theta_{(s),1}(M_{(s)}-1,1) & \Theta_{(s),1}(M_{(s)}-1,2) & \cdots & \Theta_{(s),1}(M_{(s)}-1,M_{(s)})
					\end{bmatrix}\label{le2:5}\\
					&\Theta_{(s),2}=\begin{bmatrix}
                    \Theta_{(s),2}(M_{(s)},0) & \Theta_{(s),2}(M_{(s)},1) & \cdots & \Theta_{(s),2}(M_{(s)},M_{(s)})
					\end{bmatrix}\label{le2:6}
				\end{align}
                \end{small}
				{\noindent}\rule[-10pt]{18.2cm}{0.05em}\\ 
			\end{figure*}
            \begin{small}
			\begin{align*}
				&\Theta_{(s),1}(\flat,0)=\sum_{u=0}^{\flat}p_{(s)}(u)p_{(s)}(\flat-u)\\
                &\qquad\quad+\sum_{t=1}^{\flat+1}\sum_{u=0}^{\flat+1-t}p_{(s)}(u)p_{(s)}(\flat+1-t-u)\\
				&\Theta_{(s),1}(\flat,1)=\sum_{u=0}^{\flat}p_{(s)}(u)p_{(s)}(\flat-u)\\
                &\qquad\quad+\sum_{t=2}^{\flat+2}\sum_{u=0}^{\flat+2-t}p_{(s)}(u)p_{(s)}(\flat+2-t-u)\\
				&\cdots\cdots\\
				&\Theta_{(s),1}(\flat,\flat)=\sum_{u=0}^{1}p_{(s)}(u)p_{(s)}(1-u)\\
                &\qquad\quad+\sum_{t=\flat+1}^{\flat+2}\sum_{u=0}^{\flat+2-t}p_{(s)}(u)p_{(s)}(\flat+2-t-u)\\
				&\Theta_{(s),1}(\flat,\flat+1)=2p_{(s)}^2(0)\\
                &\Theta_{(s),2}(M_{(s)},0)=\sum_{d=0}^{\infty}\sum_{w=0}^{\infty}\sum_{u=0}^{M_{(s)}+w}p_{(s)}(u+d)p_{(s)}(M_{(s)}+w-u)\nonumber\\
                &\qquad\quad+\sum_{d=0}^{\infty}\sum_{w=0}^{\infty}\sum_{t=1}^{M_{(s)}}\sum_{u=0}^{M_{(s)}+1+w-t}p_{(s)}(u+d)\nonumber\\
				&\qquad\qquad\times p_{(s)}(M_{(s)}+1+w-t-u)\\
                &\Theta_{(s),2}(M_{(s)},1)=\sum_{d=0}^{\infty}\sum_{w=0}^{\infty}\sum_{u=0}^{M_{(s)}+w}p_{(s)}(u+d)p_{(s)}(M_{(s)}+w-u)\nonumber\\
                &\qquad\quad+\sum_{d=0}^{\infty}\sum_{w=0}^{\infty}\sum_{t=2}^{M_{(s)}+1}\sum_{u=0}^{M_{(s)}+2+w-t}p_{(s)}(u+d)\nonumber\\
				&\qquad\qquad\times p_{(s)}(M_{(s)}+2+w-t-u)\\
				&\cdots\cdots\\
                &\Theta_{(s),2}(M_{(s)},M_{(s)})=2\sum_{d=0}^{\infty}\sum_{w=0}^{\infty}\sum_{u=0}^{1+w}p_{(s)}(u+d)p_{(s)}(1+w-u)
			\end{align*}
            \end{small}
		\end{lemma}
		\begin{IEEEproof}
			For $0\leq \flat < M_{(s)}$	
            \begin{small}
			\begin{align*}
				&\text{Prob}\{\varpi_{(s)}(l+1,k+1)=\flat\}\\
				&=\text{Prob}\{\varpi_{(s)}(l+1,k)+\varpi_{(s)}(l,k+1)=0,\\
				&\qquad\qquad\varsigma_{(s)}(l+1,k)+\varsigma_{(s)}(l,k+1)=\flat\}\\
				&\quad+\sum_{t=1}^{\flat+1}\text{Prob}\{\varpi_{(s)}(l+1,k)+\varpi_{(s)}(l,k+1)=t,\\
				&\qquad\qquad\varsigma_{(s)}(l+1,k)+\varsigma_{(s)}(l,k+1)=\flat+1-t\}\\
				&\quad+\sum_{t=2}^{\flat+2}\text{Prob}\{\varpi_{(s)}(l+1,k)+\varpi_{(s)}(l,k+1)=t,\\
				&\qquad\qquad\varsigma_{(s)}(l+1,k)+\varsigma_{(s)}(l,k+1)=\flat+2-t\}
			\end{align*}
            \end{small}
            The harvested energy $\varpi_{(s)}(l+1,k)+\varpi_{(s)}(l,k+1)$ is a stochastic process and it uncorrelated to $\varsigma_{(s)}(l+1,k)+\varsigma_{(s)}(l,k+1)$. Retrospecting the definition of harvested energy probability distribution and the basic formula of probability i.e., $\text{Prob}\{AB\}\leq\text{Prob}\{A\}+\text{Prob}\{B\}$.
			Thereby, the first term probability above can be rewritten as
            \begin{small}
			\begin{align*}
				&\text{Prob}\{\varpi_{(s)}(l+1,k)+\varpi_{(s)}(l,k+1)=0,\\
				&\quad\qquad\varsigma_{(s)}(l+1,k)+\varsigma_{(s)}(l,k+1)=\flat\}\\
                &=\sum_{u=0}^{\flat}p_{(s)}(u)p_{(s)}(\flat-u)\text{Prob}\{\varpi_{(s)}(l+1,k)=0,\varpi_{(s)}(l,k+1)=0\}\\
                &\leq\sum_{u=0}^{\flat}p_{(s)}(u)p_{(s)}(\flat-u)\bigl\{\text{Prob}\{\varpi_{(s)}(l+1,k)=0\}\\
				&\quad+\text{Prob}\{\varpi_{(s)}(l,k+1)=0\}\bigr\}
			\end{align*}
            \end{small}
			Simplify the second and third terms along the same lines, it is easy to obtain that
            \begin{small}
			\begin{align*}
				&\sum_{t=1}^{\flat+1}\text{Prob}\{\varpi_{(s)}(l+1,k)+\varpi_{(s)}(l,k+1)=t,\\
				&\quad\qquad\varsigma_{(s)}(l+1,k)+\varsigma_{(s)}(l,k+1)=\flat+1-t\}\\
                &\leq\sum_{t=1}^{\flat+1}\Biggl\{\sum\nolimits_{u=0}^{\flat+1-t}p_{(s)}(u)p_{(s)}(\flat+1-t-u)\\
				&\quad\times\Big[\text{Prob}\{\varpi_{(s)}(l+1,k)=0\}+\text{Prob}\{\varpi_{(s)}(l,k+1)=t\}\\
				&\quad+\text{Prob}\{\varpi_{(s)}(l+1,k)=t\}+\text{Prob}\{\varpi_{(s)}(l,k+1)=0\}\Big]\Biggr\}\\
				&\sum_{t=2}^{\flat+2}\text{Prob}\{\varpi_{(s)}(l+1,k)+\varpi_{(s)}(l,k+1)=t,\\
				&\quad\qquad\varsigma_{(s)}(l+1,k)+\varsigma_{(s)}(l,k+1)=\flat+2-t\}\\
                &\leq\sum_{t=2}^{\flat+2}\Biggl\{\sum_{u=0}^{\flat+2-t}p_{(s)}(u)p_{(s)}(\flat+2-t-u)\\
				&\quad\times\sum_{v=1}^{t-1}\Big[\text{Prob}\{\varpi_{(s)}(l+1,k)=v\}\\
				&\quad\qquad+\text{Prob}\{\varpi_{(s)}(l+1,k)=t-v\}\Big]\Biggr\}
			\end{align*}
            \end{small}
			
			Based on the above derivation, it can be inferred that
            \begin{small}
			\begin{align}\label{lem2:3}
				&\text{Prob}\{\varpi_{(s)}(l+1,k+1)=\flat\}\nonumber\\
				&\leq\Biggl\{\sum_{u=0}^{\flat}p_{(s)}(u)p_{(s)}(\flat-u)\nonumber\\
                &\quad\quad+\sum_{t=1}^{\flat+1}\sum_{u=0}^{\flat+1-t}p_{(s)}(u)p_{(s)}(\flat+1-t-u)\Biggr\}\nonumber\\
                &\quad\quad\times\Bigl\{\text{Prob}\{\varpi_{(s)}(l+1,k)=0\}+\text{Prob}\{\varpi_{(s)}(l,k+1)=0\}\Bigr\}\nonumber\\
				&\quad+\Biggl\{\sum_{u=0}^{\flat}p_{(s)}(u)p_{(s)}(\flat-u)\nonumber\\
                &\quad\quad+\sum_{t=2}^{\flat+2}\sum_{u=0}^{\flat+2-t}p_{(s)}(u)p_{(s)}(\flat+2-t-u)\Biggr\}\nonumber\\
                &\qquad\times\Bigl\{\text{Prob}\{\varpi_{(s)}(l+1,k)=1\}+\text{Prob}\{\varpi_{(s)}(l,k+1)=1\}\Bigr\}\nonumber\\
				&\quad+\cdots\cdots\nonumber\\
				&\quad+\Biggl\{\sum_{u=0}^{1}p_{(s)}(u)p_{(s)}(1-u)\nonumber\\
                &\quad\quad+\sum_{t=\flat+1}^{\flat+2}\sum_{u=0}^{\flat+2-t}p_{(s)}(u)p_{(s)}(\flat+2-t-u)\Biggr\}\nonumber\\
                &\quad\quad\times\Bigl\{\text{Prob}\{\varpi_{(s)}(l+1,k)=\flat\}+\text{Prob}\{\varpi_{(s)}(l,k+1)=\flat\}\Bigr\}\nonumber\\
                &\quad+\bigl\{p_{(s)}^2(0)+p_{(s)}^2(0)\bigr\}\Bigl\{\text{Prob}\{\varpi_{(s)}(l+1,k)=\flat+1\}\nonumber\\
				&\quad\quad+\text{Prob}\{\varpi_{(s)}(l,k+1)=\flat+1\}\Bigr\}
			\end{align}
            \end{small}
			
			For $\flat=M_{(s)}$,
            \begin{small}
			\begin{align*}
				&\text{Prob}\{\varpi_{(s)}(l+1,k+1)=M_{(s)}\}\\
                &=\sum_{d=0}^{\infty}\sum_{w=0}^{\infty}\text{Prob}\{\varpi_{(s)}(l+1,k)+\varpi_{(s)}(l,k+1)=0,\\
				&\quad\qquad\varsigma_{(s)}(l+1,k)+\varsigma_{(s)}(l,k+1)=M_{(s)}+d+w\}\\
                &\quad+\sum_{d=0}^{\infty}\sum_{w=0}^{\infty}\sum_{t=1}^{M_{(s)}}\text{Prob}\{\varpi_{(s)}(l+1,k)+\varpi_{(s)}(l,k+1)=t,\\
				&\quad\qquad\varsigma_{(s)}(l+1,k)+\varsigma_{(s)}(l,k+1)=M_{(s)}+1+d+w-t\}\\
                &\quad+\sum_{d=0}^{\infty}\sum_{w=0}^{\infty}\sum_{t=2}^{M_{(s)}+1}\text{Prob}\{\varpi_{(s)}(l+1,k)+\varpi_{(s)}(l,k+1)=t,\\
				&\quad\qquad\varsigma_{(s)}(l+1,k)+\varsigma_{(s)}(l,k+1)=M_{(s)}+2+d+w-t\}
			\end{align*}
            \end{small}
			
			Similar derivation to the first case, it can be inferred that
            \begin{small}
			\begin{align}\label{lem2:4}
				&\text{Prob}\{\varpi_{s}(l+1,k+1)=M_{(s)}\}\nonumber\\
                &\leq\Biggl\{\sum_{d=0}^{\infty}\sum_{w=0}^{\infty}\sum_{u=0}^{M_{(s)}+w}p_{(s)}(u+d)p_{(s)}(M_{(s)}+w-u)\nonumber\\
                &\quad+\sum_{d=0}^{\infty}\sum_{w=0}^{\infty}\sum_{t=1}^{M_{(s)}}\sum_{u=0}^{M_{(s)}+1+w-t}p_{(s)}(u+d)\nonumber\\
				&\qquad\times p_{(s)}(M_{(s)}+1+w-t-u)\Biggr\}\nonumber\\
                &\quad\times\biggl\{\text{Prob}\{\varsigma_{(s)}(l+1,k)=0\}+\text{Prob}\{\varsigma_{(s)}(l,k+1)=0\}\biggr\}\nonumber\\
                &+\Biggl\{\sum_{d=0}^{\infty}\sum_{w=0}^{\infty}\sum_{u=0}^{M_{(s)}+w}p_{(s)}(u+d)p_{(s)}(M_{(s)}+w-u)\nonumber\\
                &\quad+\sum_{d=0}^{\infty}\sum_{w=0}^{\infty}\sum_{t=2}^{M_{(s)}+1}\sum_{u=0}^{M_{(s)}+2+w-t}p_{(s)}(u+d)\nonumber\\
				&\quad\times p_{(s)}(M_{(s)}+2+w-t-u)\Biggr\}\nonumber\\
                &\quad\times\biggl\{\text{Prob}\{\varsigma_{(s)}(l+1,k)=1\}+\text{Prob}\{\varsigma_{(s)}(l,k+1)=1\}\biggr\}\nonumber\\
				&+\cdots\cdots\nonumber\\
                &+\Biggl\{\sum_{d=0}^{\infty}\sum_{w=0}^{\infty}\sum_{u=0}^{1+w}p_{(s)}(u+d)+p_{(s)}(1+w-u)\nonumber\\
                &\quad+\sum_{d=0}^{\infty}\sum_{w=0}^{\infty}\sum_{u=0}^{1+w}p_{(s)}(u+d)p_{(s)}(1+w-u)\Biggr\}\nonumber\\
				&\quad\times \Bigl\{\text{Prob}\{\varpi_{(s)}(l+1,k)=M_{(s)}\}\nonumber\\
				&\qquad+\text{Prob}\{\varpi_{(s)}(l,k+1)=M_{(s)}\}\Bigr\}
			\end{align}
            \end{small}
			By combining \eqref{lem2:3} and \eqref{lem2:4}, it is easy to conclude that
            \begin{small}
			\begin{align*}
				\digamma_{(s)}(l+1,k+1)\leq Q_{(s)}(l,k)\big[\digamma_{(s)}(l+1,k)+\digamma_{(s)}(l,k+1)\big]
			\end{align*}
            \end{small}
			The following proves this lemma by using mathematical induction.
			
            {\it Initial Step:} Retrospecting the initial conditions, we have \eqref{lem2:1} is hold for $(l,k)=(1,1)$.
			
            {\it Inductive Step:} Let the assertion \eqref{lem2:1} is true for $(l,k)\in\{\\ (l_0,k_0)|l_0\in[1,\varkappa],k_0\in[1,\varkappa];l_0+k_0=z\}$ with $z\in[2,2\varkappa-1]$. Then, we next need to verify that \eqref{lem2:1} is true for $(l,k)\in\{(l_0,k_0)|l_0\in[1,i_0],k_0\in[1,j_0];l_0+k_0=z+1\}$.
            \begin{small}
			\begin{align*}
				&\mathcal{F}_{(s)}(l,k)-\digamma_{(s)}(l,k)\\
				&\quad\geq Q_{(s)}\big[\mathcal{F}_{(s)}(l,k-1)-\digamma_{(s)}(l,k-1)\\
				&\qquad+\mathcal{F}_{(s)}(l-1,k)-\digamma_{(s)}(l-1,k)\big]\\
				&\quad\geq0
			\end{align*}
            \end{small}
			This completes the proof of this lemma.
		\end{IEEEproof}	
        \begin{remark}
            In view of the bidirectional evolution characteristics of 2-D systems, it is difficult to obtain the probability distribution of energy level directly. In light to the above lemma, find an upper bound of the probability distribution of the energy level $\varpi_{(s)}(l,k)$. The activated probability can be approximately calculated as
			\begin{align*}
                &\varrho_{(s)}(l,k)=\mathcal{E}[\mathbf{1}_{\{\breve{\varpi}_{(s)}(l,k)>0\}}]=[0 \quad \mathbf{1}_{M_{(s)}}]\mathcal{F}_{(s)}(l,k)\\ &\tilde{\varrho}_{(s)}(l,k)=Var[\mathbf{1}_{\{\breve{\varpi}_{(s)}(l,k)>0\}}]=\varrho_{(s)}(l,k)(1-\varrho_{(s)}(l,k)).
			\end{align*}
		\end{remark}
    \subsection{Design of the estimation gains}
        In this subsection, the design details of the estimation gains of the $\hbar$+1-step recursive filter of 2-D systems \eqref{CW2a:1}-\eqref{CW2a:2} will be given.
		
		For the convenience of subsequent narration, define
		\begin{align}
			&X(l,k)\triangleq \mathcal{E}\{x(l,k)x^\mathrm{T}(l,k)\}\label{CW3d:1}\\
            &\Xi_p^{\gamma}(l,k)\triangleq \mathcal{E}\{\tilde{x}_p^{\gamma}(l,k)(\tilde{x}_p^{\gamma}(l,k))^\mathrm{T}\}\label{CW3d:2}\\
            &\Xi_u^{\tau}(l,k)\triangleq \mathcal{E}\{\tilde{x}_u^{\gamma}(l,k)(\tilde{x}_u^{\gamma}(l,k))^\mathrm{T}\}\label{CW3d:3}
		\end{align}
		where $\gamma\in[1,\hbar+1]$.
		
        Before presenting the main conclusion in this subsection, the unbiased proof of the proposed $\hbar$+1-step recursive filter is given by mathematical induction.
		\begin{lemma}\label{lem3}
            Consider 2-D systems \eqref{CW2a:1} and \eqref{CW2a:2}, the proposed $\hbar$+1-step recursive filter is unbiased, that is to say the errors \eqref{CW3b:6}-\eqref{CW3b:8} are zero-mean variables.
		\end{lemma}
		\begin{IEEEproof}
			The proof of this lemma is given by mathematical induction. According to the initial conditions
            $\hat{x}_u^1(l,0)=\hat{x}_u(l,0)=\mathcal{E}\{x(l,0)\}$ and $\hat{x}_u^1(0,k)=\hat{x}_u(0,k)=\mathcal{E}\{x(0,k)\}$. Thus, one has
			\begin{align*}
				\mathcal{E}\{\tilde{x}_u^1(l,0)\}&=\mathcal{E}\{\tilde{x}_u^1(0,k)\}=0\\
				&(0\leq l \leq i_1,0\leq k \leq j_1)
			\end{align*}
			In light of the statistical properties of the $g(x(l,k),\xi(l,k))$ and the random variable $w(l,k)$,
			\begin{align*}
				\mathcal{E}\{\tilde{x}_p^1(1,1)\}&=A_1(1,0)\mathcal{E}\{\tilde{x}_u^1(1,0)\}\\
				&\quad+A_2(0,1)\mathcal{E}\{\tilde{x}_u^1(0,1)\}\\
				&=0
			\end{align*}
            In view of the $\mathbf{1}_{\varpi_{\hbar}(l,k)>0}-\varrho_{\hbar}(l,k)$, $v_{\hbar}(l,k)$, and $h_{\hbar}(0,\\ \gamma(l,k))$ are zero-mean. It is easy to know that $\mathcal{E}\{\tilde{x}_u^1(1,1)\}=0$. Next, assume that $\mathcal{E}\{\tilde{x}_u^1(l_0,1)\}=0$ and $\mathcal{E}\{\tilde{x}_u^1(1,k_0)\}=0$ are valid for given scalars $2\leq l_0 \leq i_0-1, 2\leq k_0\leq j_0-1$.
			The following equations are hold from \eqref{CW3b:6}
			\begin{align*}
				\mathcal{E}\{\tilde{x}_p^1(l_0+1,1)\}&=0, \mathcal{E}\{\tilde{x}_p^1(1,k_0+1)\}=0\\
				\mathcal{E}\{\tilde{x}_u^1(l_0+1,1)\}&=0, \mathcal{E}\{ \tilde{x}_u^1(1,k_0+1)\}=0
			\end{align*}
            Based on the induction assumption, it can be inferred that the following qualities are hold for $(l,k)\in\{(l_0,1)|2\leq l_0 \leq i_0\}\cup \{(1,k_0)|2\leq k_0 \leq j_0\}$,
			\begin{align}
				\mathcal{E}\{\tilde{x}_p^1(l,k)\}=\mathcal{E}\{\tilde{x}_u^1(l,k)\}=0\label{CW3d:4}
			\end{align}
            Further prove that \eqref{CW3d:4} is valid for $(l,k)\in\{(r,k_0)|l_0\leq r \leq i_0\}\cup \{(l_0,s)|k_0\leq s \leq j_0\}$ with given constants $l_0,k_0 (2\leq l_0\leq i_0-1, 2\leq k_0 \leq j_0-1)$, it is easy to know that
		 	\begin{align*}
		 		\mathcal{E}\{\tilde{x}_p^1(l_0+1,k_0+1)\}=0
		 	\end{align*}
	 		In view of the induction assumption again, one has
	 		\begin{align*}
	 			&\mathcal{E}\{\tilde{x}_p^1(l_0+1,k)\}=\mathcal{E}\{\tilde{x}_u^1(l_0+1,k)\}=0\\
	 			&\mathcal{E}\{\tilde{x}_p^1(l,k_0+1)\}=\mathcal{E}\{\tilde{x}_u^1(l,k_0+1)\}=0
	 		\end{align*}
            for all $l(l_0+1\leq l \leq i_0)$ and $k(k_0+1\leq k \leq j_0)$. Then repeat this process as $l$ and $k$ increase, respectively, it can be proven that \eqref{CW3d:4} hold for all $(l,k)\in N_0$.
		\end{IEEEproof}
		
        The following lemmas give the updated terms of the second-order moment $X(l,k)$, which will play an irreplaceable role in the recursive calculation of the error covariance.
		\begin{lemma}\label{lem4}
            Consider the 2-D systems \eqref{CW2a:1} and \eqref{CW2a:2}, with the initial conditions $X(l,0)=\Xi(l,0)$ and $X(0,k)=\Xi(0,k)$, the recursive evolution is as follow:
            \begin{small}
			\begin{align}\label{CW3d:5}
				X(l,k)=&A_1(l,k-1)X(l,k-1)A_1^\mathrm{T}(l,k-1)\nonumber\\
				&+A_2(l-1,k)X(l-1,k)A_2^\mathrm{T}(l-1,k)\nonumber\\
				&+B_1(l,k-1)Q(l,k-1)B_1^\mathrm{T}(l,k-1)\nonumber\\
				&+B_2(l-1,k)Q(l-1,k)B_2^\mathrm{T}(l-1,k)\nonumber\\
				&+\sum_{\mu=1}^{r}\Delta_{g,\mu}tr\{(X(l,k-1)+X(l-1,k))\Sigma_{g,\mu}\}\nonumber\\
				&+A_1(l,k-1)\mathcal{E}\{x(l,k-1)x^\mathrm{T}(l-1,k)\}\nonumber\\
				&\times A_2^\mathrm{T}(l-1,k)+A_2(l-1,k)\nonumber\\
				&\times \mathcal{E}\{x(l-1,k)x^\mathrm{T}(l,k-1)\}A_1^\mathrm{T}(l,k-1)
			\end{align}
            \end{small}
		\end{lemma}
		\begin{IEEEproof}
            Retrospecting the statistical properties of variables $w(l,k)$ and stochastic nonlinearity $g(x(l,k),\xi(l,k))$, we have
            \begin{small}
			\begin{align*}
				\mathcal{E}&\{x(l,k)g^\mathrm{T}(x(u,v),\xi(u,v))\}\\
				&=\mathcal{E}\{\mathcal{E}\{x(l,k)g^\mathrm{T}(x(u,v),\xi(u,v))|x(l,k)\}\}\\
				&=\mathcal{E}\{x(l,k)\mathcal{E}\{g(x(u,v),\xi(u,v))|x(l,k)\}\}\\
				&=0\quad (u,v)\in\{(u_1,v_1)|u_1>l\ \text{or}\ v_1>k\}\\
				\mathcal{E}&\{g(x(l,k),\xi(l,k))g^\mathrm{T}(x(u,v),\xi(u,v))\}\\
				&=\mathcal{E}\{\mathcal{E}\{g(x(l,k),\xi(l,k))g^\mathrm{T}(x(u,v),\xi(u,v))\}|x(l,k)\}\\
				&=\sum_{\mu=1}^{r}\Delta_{g,\mu}\mathcal{E}\{x^\mathrm{T}(l,k)\Sigma_{g,\mu}x(l,k)\}\delta(l,u)\delta(k,v)\\
				&=\sum_{\mu=1}^{r}\Delta_{g,\mu}tr\{X(l,k)\Sigma_{g,\mu}\}\delta(l,u)\delta(k,v)
			\end{align*}
            \end{small}
			Then, this lemma is proven from \eqref{CW2a:1}.
		\end{IEEEproof}
        \begin{lemma}
            Consider the 2-D systems \eqref{CW2a:1} and \eqref{CW2a:2}, the evolution of the prediction error covariance are given as follows:
			
            $\mathbf{Step\  1}$: For $(l,k)\in \mathcal{N}_{0}$, with the initial conditions $\Xi_p^1(l,0)\\ =\Xi(l,0)$ and $\Xi_p^1(0,k)=\Xi(0,k)$, the one-step prediction error covariance is calculated as
            \begin{small}
			\begin{align}\label{CW3d:6}
				\Xi_p^1(l,k)=&A_1(l,k-1)\Xi_u^1(l,k-1)A_1^\mathrm{T}(l,k-1)\nonumber\\
				&+A_2(l-1,k)\Xi_u^1(l-1,k)A_2^\mathrm{T}(l-1,k)\nonumber\\
				&+B_1(l,k-1)Q(l,k-1)B_1^\mathrm{T}(l,k-1)\nonumber\\
				&+B_2(l-1,k)Q(l-1,k)B_2^\mathrm{T}(l-1,k)\nonumber\\
				&+\sum_{\mu=1}^{r}\Delta_{g,\mu}tr\{(X(l,k-1)+X(l-1,k))\Sigma_{g,\mu}\}\nonumber\\
				&+A_1(l,k-1)\mathcal{E}\{\tilde{x}_u^1(l,k-1)(\tilde{x}_u^1(l-1,k))^\mathrm{T}\}\nonumber\\
				&\times A_2^\mathrm{T}(l-1,k)+A_2(l-1,k)\nonumber\\
				&\times\mathcal{E}\{\tilde{x}_u^1(l-1,k)(\tilde{x}_u^1(l,k-1))^\mathrm{T}\}A_1^\mathrm{T}(l,k-1)
			\end{align}
            \end{small}
		
            $\mathbf{Step\  2}$: For $(l,k)\in \mathcal{N}_{1}$, with the initial conditions $\Xi_p^2(l,0)=\Xi(l,0)(i_0+1\leq l \leq i_1), \Xi_p^2(l,j_0)=\Xi_p^1(l,j_0)(0\leq l \leq i_0)$ and $\Xi_p^2(0,k)=\Xi(0,k)(j_0+1\leq k \leq j_1), \Xi_p^2(i_0,k)=\Xi_p^1(i_0,k)(0\leq k \leq j_0)$, the one-step prediction error covariance is calculated as
            \begin{small}
			\begin{align}\label{CW3d:7}
				\Xi_p^2(l,k)=&A_1(l,k-1)\Xi_u^2(l,k-1)A_1^\mathrm{T}(l,k-1)\nonumber\\
				&+A_2(l-1,k)\Xi_u^2(l-1,k)A_2^\mathrm{T}(l-1,k)\nonumber\\
				&+B_1(l,k-1)Q(l,k-1)B_1^\mathrm{T}(l,k-1)\nonumber\\
				&+B_2(l-1,k)Q(l-1,k)B_2^\mathrm{T}(l-1,k)\nonumber\\
				&+\sum_{\mu=1}^{r}\Delta_{g,\mu}tr\{(X(l,k-1)+X(l-1,k))\Sigma_{g,\mu}\}\nonumber\\
				&+A_1(l,k-1)\mathcal{E}\{\tilde{x}_u^2(l,k-1)(\tilde{x}_u^2(l-1,k))^\mathrm{T}\}\nonumber\\
				&\times A_2^\mathrm{T}(l-1,k)+A_2(l-1,k)\nonumber\\
				&\times\mathcal{E}\{\tilde{x}_u^2(l-1,k)(\tilde{x}_u^2(l,k-1))^\mathrm{T}\}A_1^\mathrm{T}(l,k-1)
			\end{align}
            \end{small}
			
			$\qquad\qquad\qquad\qquad\qquad \vdots$
			
            $\mathbf{Step\  \hbar+1}$: For $(l,k)\in \mathcal{N}_{\hbar}$, with the initial conditions $\Xi_p^{\hbar+1}(l,0)=\Xi(l,0)(i_{\hbar-1}+1\leq l \leq i_{\hbar}), \Xi_p^{\hbar+1}(l,j_{\hbar-1})=\Xi_p^{\hbar}(l,j_{\hbar-1})(0\leq l \leq i_{\hbar-1})$ and $\Xi_p^{\hbar+1}(0,k)=\Xi(0,k)(j_{\hbar-1}+1\leq k \leq j_{\hbar}), \Xi_p^{\hbar+1}(i_{\hbar-1},k)=\Xi_p^{\hbar}(i_{\hbar-1},k)(0\leq k \leq j_{\hbar-1})$, the one-step prediction error covariance is calculated as
            \begin{small}
			\begin{align}\label{CW3d:8}
				\Xi_p^{\hbar+1}(l,k)=&A_1(l,k-1)\Xi_u^{\hbar+1}(l,k-1)A_1^\mathrm{T}(l,k-1)\nonumber\\
				&+A_2(l-1,k)\Xi_u^{\hbar+1}(l-1,k)A_2^\mathrm{T}(l-1,k)\nonumber\\
				&+B_1(l,k-1)Q(l,k-1)B_1^\mathrm{T}(l,k-1)\nonumber\\
				&+B_2(l-1,k)Q(l-1,k)B_2^\mathrm{T}(l-1,k)\nonumber\\
				&+\sum_{\mu=1}^{r}\Delta_{g,\mu}tr\{(X(l,k-1)+X(l-1,k))\Sigma_{g,\mu}\}\nonumber\\
				&+A_1(l,k-1)\mathcal{E}\{\tilde{x}_u^{\hbar+1}(l,k-1)(\tilde{x}_u^{\hbar+1}(l-1,k))^\mathrm{T}\}\nonumber\\
				&\times A_2^\mathrm{T}(l-1,k)+A_2(l-1,k)\nonumber\\
                &\times\mathcal{E}\{\tilde{x}_u^{\hbar+1}(l-1,k)(\tilde{x}_u^{\hbar+1}(l,k-1))^\mathrm{T}\}A_1^\mathrm{T}(l,k-1)
			\end{align}
            \end{small}
		\end{lemma}
		\begin{IEEEproof}
            It follows from \eqref{CW3b:1a}-\eqref{CW3b:3a}, retrospecting the statistical properties of variables $w(l,k)$ and stochastic nonlinearity $g(x(l,k),\xi(l,k))$. Similar to the proof of Lemma~\ref{lem4}, we will not elaborate on it here.
		\end{IEEEproof}
		\begin{theorem}\label{th1}
            Consider the 2-D systems \eqref{CW2a:1} and \eqref{CW2a:2}, the evolution of the minimum filtering error covariance and the estimation gains are given below,
			
				\begin{subequations}\label{CW3d:9}
				$\mathbf{Step\  1}$: For $(l,k)\in \mathcal{N}_{0}$
				\begin{align}
                    K_1(l,k)=&\Xi_p^1(l,k)C_{\hbar}^\mathrm{T}(l,k)\varrho_{\hbar}^\mathrm{T}(l,k)\bar{R}_{\hbar}^{-1}(l,k)\label{CW3d:9a}\\
					\Xi_u^1(l,k)=&\Xi_p^1(l,k)-K_1(l,k)\varrho_{\hbar}(l,k)\nonumber\\
					&\times C_{\hbar}(l,k)\Xi_p^1(l,k)\label{CW3d:9c}
				\end{align}
			\end{subequations}
		
			\begin{subequations}\label{CW3d:10}
				$\mathbf{Step\  2}$: For $(l,k)\in \mathcal{N}_{1}$
				\begin{align}
                    K_2(l,k)=&\Xi_p^2(l,k)C_{\hbar-1}^\mathrm{T}(l,k)\varrho_{\hbar-1}^\mathrm{T}(l,k)\bar{R}_{\hbar-1}^{-1}(l,k)\label{CW3d:10a}\\
					\Xi_u^2(l,k)=&\Xi_p^2(l,k)-K_2(l,k)\varrho_{\hbar-1}(l,k)\nonumber\\
					&\times C_{\hbar-1}(l,k)\Xi_p^2(l,k)\label{CW3d:10c}
				\end{align}
			\end{subequations}
			
			$\qquad\qquad\qquad\qquad\qquad \vdots$
			
			\begin{subequations}\label{CW3d:11}
				$\mathbf{Step\  \hbar+1}$: For $(l,k)\in \mathcal{N}_{\hbar}$
				\begin{align}
                    K_{\hbar+1}(l,k)=&\Xi_p^{\hbar+1}(l,k)C_{0}^\mathrm{T}(l,k)\varrho_{0}^\mathrm{T}(l,k)\bar{R}_{0}^{-1}(l,k)\label{CW3d:11a}\\
					\Xi_u^{\hbar+1}(l,k)=&\Xi_p^{\hbar+1}(l,k)-K_{\hbar+1}(l,k)\varrho_{0}(l,k)\nonumber\\
					&\times C_{0}(l,k)\Xi_p^{\hbar+1}(l,k)\label{CW3d:11c}
				\end{align}
			\end{subequations}
            where
            \begin{small}
            \begin{align*}
                \bar{R}_{s}(l,k)\triangleq&\varrho_{s}(l,k)\big(C_{s}(l,k)\Xi_p^{\hbar+1-s}(l,k)C_{s}^\mathrm{T}(l,k)\nonumber\\
                &+R_{s}(l,k)+\sum_{\mu=1}^{r}\Delta_{h_s,\mu}tr\{X(l,k)\Sigma_{h_{s},\mu}\}\big)\varrho_{s}(l,k)\\
                &+\tilde{\varrho}_{s}(l,k)\circ(C_{s}(l,k)X(l,k)C_{s}^\mathrm{T}(l,k))
            \end{align*}
            \end{small}
		\end{theorem}
		\begin{IEEEproof}
            Reviewing the statistical property of variable $v_{s}(l,k)$, it is easy to conclude that $v_s(l,k)$ is uncorrelated with $x(l,k)$ and $\tilde{x}_u^{\hbar+1-s}(l,k)$. In accordance with the projection theorem, it can be seen that $\mathbf{1}_{\varpi_{s}(l,k)>0}-\varrho_s(l,k)$ and $\varrho_s(l,k)$ are independent to each other. Inserting \eqref{CW3b:6b} into \eqref{CW3d:1} yields
			\begin{small}
            \begin{align*}
				\Xi_u^1(l,k)=&[I-K_1(l,k)\varrho_{\hbar}(l,k)C_{\hbar}(l,k)]\Xi_p^1(l,k)\\
				&\times [I-K_1(l,k)\varrho_{\hbar}(l,k)C_{\hbar}(l,k)]^\mathrm{T}\\
				&+K_1(l,k)\Big\{\tilde{\varrho}_{\hbar}(l,k)\circ(C_{\hbar}(l,k)X(l,k)C_{\hbar}^\mathrm{T}(l,k))\\
                &+\varrho_{\hbar}(l,k)\big(R_{\hbar}(l,k)+\sum_{\mu=1}^{r}\Delta_{h_\hbar,\mu}tr\{X(l,k)\Sigma_{h_{\hbar},\mu}\}\big)\\
                &\times\varrho_{\hbar}(l,k)\Big\}K_1^\mathrm{T}(l,k)
			\end{align*}
            \end{small}
			By applying the square completion method to the above formula, it can be obtained that
			\begin{small}
            \begin{align}\label{CW3d:12}
				\Xi_u^1(l,k)=&\Xi_p^1(l,k)+\big(K_1(l,k)-\mathcal{K}_1(l,k)\big)\nonumber\\
				&\times \bar{R}_{\hbar}(l,k)\big(K_1(l,k)-\mathcal{K}_1(l,k)\big)^\mathrm{T}\nonumber\\
				&-\mathcal{K}_1(l,k)\bar{R}_{\hbar}(l,k)\mathcal{K}_1^\mathrm{T}(l,k)
			\end{align}
            \end{small}
			where
            $\mathcal{K}_1(l,k)=\Xi_p^1(l,k)C_{\hbar}^\mathrm{T}(l,k)\varrho_{\hbar}^\mathrm{T}\bar{R}_{\hbar}^{-1}(l,k)$, it is easy to conclude that $\Xi_u^1(l,k)$ is minimized if and only if the estimation gain $K_1(l,k)=\mathcal{K}_1(l,k)$. Substituting \eqref{CW3d:9a} into \eqref{CW3d:12} yields
			\begin{align*}
				\Xi_u^1(l,k)&=\Xi_p^1(l,k)-\mathcal{K}_1(l,k)\bar{R}_{\hbar}(l,k)\mathcal{K}_1^\mathrm{T}(l,k)\\
				&=\Xi_p^1(l,k)-K_1(l,k)\varrho_\hbar(l,k)C_\hbar(l,k)\Xi_p^1(l,k)
			\end{align*}
            This completes the proof.
		\end{IEEEproof}

        The above estimation gains minimizes the filtering error covariance. Below, we will construct an upper bound in the sense of spectral norm to constrain the filtering error covariance.

\section{Performance analysis}\label{sec:4}
        In this section, let us consider the performance of the proposed filter. First, under the parameter assumption, the boundedness of the filtering error covariance is given. Then, the monotonicity analysis of the activated probability is given.
		\subsection{Boundedness analysis of the filtering error covariance}
			Before presenting the main results of this subsection, the following parameter assumption is made.
			\begin{assumption}\label{ass4}
                There exist real scalars $\overline{a}_{\iota},\overline{c}_{\gamma},\overline{q}_{\iota},\underline{q}_{\iota},\overline{r}_{\gamma},\underline{r}_{\gamma},\\ \overline{x},\overline{\delta}_{g,\mu}$, and $\overline{\sigma}_{g,\mu}$, the following inequalities satisfies for every $l,k\in[0,\varkappa],\gamma\in[0,\hbar],\mu\in[1,r]$, and $\iota\in[1,2]$:
				\begin{align*}
                    &\underline{q}_{\iota}I\leq Q_{\iota}(l,k)\leq \overline{q}_{\iota}I,\quad\quad \Delta_{g,\mu}\Delta_{g,\mu}^\mathrm{T}\leq \overline{\delta}_{g,\mu}I\\
                    &\underline{r}_{\gamma}I\leq R_{\gamma}(l,k) \leq \overline{r}_{\gamma}I,\quad\  \ C_{\tau}(l,k)C_{\gamma}^\mathrm{T}(l,k)\leq \overline{c}_{\gamma}I\\
                    &A_{\iota}(l,k)A_{\iota}^\mathrm{T}(l,k)\leq \overline{a}_{\iota}I,\quad \ tr\{X(l,k)\}\leq \overline{x} \\
                    &(\Sigma_{g,\mu}+\Sigma_{g,\mu}^\mathrm{T})(\Sigma_{g,\mu}+\Sigma_{g,\mu}^\mathrm{T})^\mathrm{T}\leq \overline{\sigma}_{g,\mu}I
				\end{align*}
				where $Q_{\iota}(l,k)=B_{\iota}(l,k)Q(l,k)B_{\iota}^\mathrm{T}(l,k)$.
			\end{assumption}
			\begin{theorem}\label{th2}
                Under Assumption~\ref{ass4}, the filtering error covariance $\Xi_u^\gamma(l,k)$ satisfy the following inequality
				\begin{align*}
					\Xi_u^\gamma(l,k)\geq\underline{\epsilon} I
				\end{align*}
				for every $l,k\in[0,\varkappa]$ with the lower bound $\underline{\epsilon}$ given as follows
				\begin{align*}
                    \underline{\epsilon}=\Big(\frac{1}{\underline{q}_1+\underline{q}_2}+\frac{\overline{c}_{\hbar+1-\gamma}}{\underline{r}_{\hbar+1-\gamma}}\Big)^{-1}
				\end{align*}
			\end{theorem}
			\begin{IEEEproof}
				Retrospecting Assumption~\ref{ass4} and \eqref{CW3d:6}, we have
                \begin{small}
				\begin{align*}
					&\Xi_p^\gamma(l,k) \geq Q_1(l,k-1)+Q_2(l-1,k)\geq(\underline{q}_1+\underline{q}_2)I\\
                    &C_\gamma^\mathrm{T}(l,k)\Big(\tilde{\varrho}_{\gamma}(l,k)\circ(C_{\gamma}(l,k)X(l,k)C_{\gamma}^\mathrm{T}(l,k))\\
                    &\quad+\varrho_{\gamma}(l,k)\big(R_{\gamma}(l,k)+\sum_{\mu=1}^{r}\Delta_{h_\gamma,\mu}tr\{X(l,k)\Sigma_{h_\gamma,\mu}\}\big)\\
					&\quad\times\varrho_{\gamma}(l,k)\Big)^{-1}C_\gamma(l,k)\leq\overline{c}_{\gamma}/\underline{r}_{\gamma}I
				\end{align*}
                \end{small}
				In view of matrix inversion formula and \eqref{CW3d:12}, it can be seen that
                \begin{small}
				\begin{align*}
                    (\Xi_u^\gamma(l,k))^{-1}&=(\Xi_p^\gamma(l,k)-K_\gamma(l,k)\bar{R}_{\hbar+1-\gamma}(l,k)K_\gamma^\mathrm{T}(l,k))^{-1}\\
                    &=C_{\hbar+1-\gamma}^\mathrm{T}(l,k)\Big(\tilde{\varrho}_{\hbar+1-\gamma}(l,k)\circ(C_{\hbar+1-\gamma}(l,k)X(l,k)\\
                    &\quad \times C_{\hbar+1-\gamma}^\mathrm{T}(l,k))+\varrho_{\gamma}(l,k)\big( R_{{\hbar+1-\gamma}}(l,k)\\
                    &\quad+\sum_{\mu=1}^{r}\Delta_{h_{\hbar+1-\gamma},\mu}tr\{X(l,k)\Sigma_{h_{\hbar+1-\gamma},\mu}\} \big)\\
					&\quad \times \varrho_{\gamma}(l,k)\Big)^{-1}C_{\hbar+1-\gamma}(l,k)+(\Xi_p^\gamma(l,k))^{-1}\\
                    &\leq \Big(\frac{1}{\underline{q}_1+\underline{q}_2}+\frac{\overline{c}_{\hbar+1-\gamma}}{\underline{r}_{\hbar+1-\gamma}}\Big) I
				\end{align*}
                \end{small}
				which implies that $\Xi_u^\gamma(l,k)\geq\underline{\epsilon} I$ for every $l,k\in[0,\varkappa]$.
			\end{IEEEproof}
		
            Theorem~\ref{th2} provide a uniform lower bound for the filtering error covariance. Next, a recursive upper bound function for the filtering error covariance will be given.
			\begin{theorem}\label{th3}
				Let Assumption~\ref{ass4} hold, the filtering error covariance satisfy the following inequality
				
				For $(l,k)\in \mathcal{N}_0^{\circ}$,
                \begin{small}
				\begin{align}\label{CW4a:1}
					\Xi_u^1(l,k)&\leq \sum_{s=1}^{l}\gamma_1\daleth(l-s,k-1)	\Xi_u^1(s,0)\nonumber\\
					&\ +\sum_{t=1}^{k}\gamma_2\daleth(l-1,k-t)	\Xi_u^1(0,t)\nonumber\\
					&\ +\sum_{s=0}^{l-1}\sum_{t=0}^{k-1}\daleth(l-s-1,k-t-1)\gimel_0I
				\end{align}
                \end{small}
				
				$\qquad\qquad\qquad\qquad\qquad \vdots$
				
				For $(l,k)\in \mathcal{N}_{\hbar}^{\circ}$,
                \begin{small}
				\begin{align}\label{CW4a:2}
                    \Xi_u^{\hbar+1}(l,k)&\leq\sum_{s=i_{\hbar-1}+1}^{l}\gamma_1\daleth(l-s,k-1)	\Xi_u^{\hbar+1}(s,j_{\hbar-1})\nonumber\\
                    &\ +\sum_{t=j_{\hbar-1}+1}^{k}\gamma_2\daleth(l-1,k-t)	\Xi_u^{\hbar+1}(i_{\hbar-1},t)\nonumber\\
					&\ +\sum_{s=i_{\hbar-1}}^{l-1}\sum_{t=i_{\hbar-1}}^{k-1}\daleth(l-s-1,k-t-1)\gimel_0I
				\end{align}
                \end{small}
                where $\gamma>0$ is a scalar, $\upsilon_1=1+\gamma,\upsilon_2=1+\gamma^{-1},\gamma_1=\upsilon_1\overline{a}_1^2,\gamma_2=\upsilon_2\overline{a}_2^2,\gimel_0=\sum_{\mu}^{r}\overline{x}\overline{\delta}_{g,\mu}\overline{\sigma}_{g,\mu}+(\overline{q}_1+\overline{q}_2)I,\daleth(\cdot,\cdot)$ with $\daleth(0,0)=\cdots=\daleth(i_{\hbar-1},j_{\hbar-1})=1$ is can be recursively calculated by
				\begin{small}
                \begin{align*}
					\daleth(0,k)&=\gamma_1\daleth(0,k-1)\\
					\daleth(l,0)&=\gamma_2\daleth(l-1,0)\\
					\daleth(l,k)&=\gamma_1\daleth(l,k-1)+\gamma_2\daleth(l-1,k)
				\end{align*}
                \end{small}
			\end{theorem}
			\begin{IEEEproof}
				For every scalar $\gamma$,
				\begin{align}\label{CW4a:3}
					\Xi_p^1(l,k)&\leq \upsilon_1A_1(l,k-1)\Xi_p^1(l,k-1)A_1^\mathrm{T}(l,k-1)\nonumber\\
					&\ +\upsilon_2A_2(l-1,k)\Xi_p^1(l-1,k)A_2^\mathrm{T}(l-1,k)\nonumber\\
                    &\ +\sum_{\mu=1}^{r}\Delta_{g,\mu}tr\{(X(l,k-1)+X(l-1,k))\Sigma_{g,\mu}\}\nonumber\\
					&\ +Q_1(l,k-1)+Q_2(l-1,k)
				\end{align}
				According to the Assumption~\ref{ass4}, we have
				\begin{align}\label{CW4a:4}
					&tr\{(X(l,k-1)+X(l-1,k))\Sigma_{g,\mu}\}\nonumber\\
                    &\quad =\frac{1}{2}\mathcal{E}\{x^\mathrm{T}(l,k-1)(\Sigma_{g,\mu}+\Sigma_{g,\mu}^\mathrm{T})x(l,k-1)\nonumber\\
					&\quad\ +x^\mathrm{T}(l-1,k)(\Sigma_{g,\mu}+\Sigma_{g,\mu}^\mathrm{T})x(l-1,k)\}\nonumber\\
					&\quad\leq\overline{x}\overline{\sigma}_{g,\mu}
				\end{align}
					Furthermore, by inserting \eqref{CW4a:4} into \eqref{CW4a:1}
					\begin{align}\label{CW4a:5}
						\Xi_p^1(l,k)\leq\gamma_1\Xi_u^1(l,k-1)+\gamma_2\Xi_u^1(l-1,k)+\gimel_0I
					\end{align}
				
				For $(l,k)\in\mathcal{N}_0^\circ$, prove the theorem by using mathematical induction.
				
                {\it Initial step}: Visibly, by employing \eqref{CW3d:9c} and \eqref{CW4a:2}, we have \eqref{CW4a:1} is hold for $(l,k)=(1,1)$.
				
                {\it Inductive step}: We assume that the assertion \eqref{CW4a:1} is true for $(l,k)\in\{(l_0,k_0)|l_0\in[1,i_0],k_0\in[1,j_0];l_0+k_0=\ell\}$ with $\ell\in[2,i_0+j_0-1]$. Then, we aim to check if the same assertion \eqref{CW4a:1} is true for $(l,k)\in\{(l_0,k_0)|l_0\in[1,i_0],k_0\in[1,j_0];l_0+k_0=\ell+1\}$.
                \begin{small}
				\begin{align*}
					&\Xi_u^1(l,k)\\
					&\ \leq \gamma_1\Xi_u^1(l,k-1)+\gamma_2\Xi_u^1(l-1,k)\\
					&\ \leq \gamma_1\Bigg\{\sum\nolimits_{s=1}^{l}\gamma_1\daleth(l-s,k-2)\Xi_u^1(s,0)\\
					&\quad +\sum_{t=1}^{k-1}\gamma_2\daleth(l-1,k-t-1)\Xi_u^1(0,t)\\
					&\quad +\sum_{s=0}^{l-1}\sum\nolimits_{t=0}^{k-2}\daleth(l-s-1,k-t-2)\gimel_0I\Bigg\}\\
					&\quad +\gamma_2\Bigg\{\sum\nolimits_{s=1}^{l-1}\gamma_1\daleth(l-s-1,k-1)\Xi_u^1(s,0)\\
					&\quad +\sum_{t=1}^{k}\gamma_2\daleth(l-2,k-t)\Xi_u^1(0,t)\\
                    &\quad +\sum_{s=0}^{l-2}\sum_{t=0}^{k-1}\daleth(l-s-2,j-t-1)\gimel_0I\Bigg\}+\gimel_0I\\
					&\ =\gamma_1^2\daleth(0,k-2)\Xi_u^1(l,0)+\gamma_2^2\daleth(l-2,0)\Xi_u^1(0,k)\\
					&\quad +\sum_{s=1}^{l-1}\Big\{\gamma_1^2\daleth(l-s,k-2)\\
					&\quad +\gamma_1\gamma_2\daleth(l-s-1,k-1)\Big\}\Xi_u^1(s,0)\\
					&\quad +\sum_{t=1}^{k-1}\Big\{\gamma_2^2\daleth(l-2,k-t)\\
					&\quad +\gamma_1\gamma_2\daleth(l-1,k-t-1)\Big\}\Xi_u^1(0,t)+\gimel_0I\\
                    &\quad +\sum_{s=0}^{l-1}\sum\nolimits_{t=0}^{k-2}\daleth(l-s-1,k-t-2)\gamma_1\gimel_0I\\
                    &\quad +\sum_{s=0}^{l-2}\sum\nolimits_{t=0}^{k-1}\daleth(l-s-2,k-t-1)\gamma_2\gimel_0I\\
					&\ =\sum_{s=1}^{l}\gamma_1\daleth(l-s,k-1)\Xi_u^1(s,0)\\
					&\quad +\sum_{t=1}^{k}\gamma_2\daleth(l-1,k-t)\Xi_u^1(0,t)\\
					&\quad +\sum_{s=0}^{l-1}\sum_{t=0}^{k-1}\daleth(l-s-1,k-t-1)\gimel_0I
				\end{align*}
                \end{small}
                Hence, by induction, we can conclude that the assertion \eqref{CW4a:1} is true for $(l,k)\in\mathcal{N}_0^\circ$. This completes the proof.
			\end{IEEEproof}
			\begin{remark}
                So far, we have derived an upper bound function for the error covariance for the developed $\hbar$+1 recursive filtering algorithm. It can be seen from the recursive structure of the upper bound function that almost all important factors affecting the complexity of the system are reflected in the main results. These factors include:
				\begin{itemize}
					\item[1)] the shift-varying system parameters;
					\item[2)] the random noise statistical property; and
					\item[3)] the initial condition of the system.
				\end{itemize}
			\end{remark}
			\begin{remark}
                It is noteworthy that, the impact of energy harvesting constraints on the filtering technique proposed in this paper is not reflected in the upper bound function of the error covariance matrix. This is due to the essential difficulty caused by the bidirectional evolution of the 2-D system in the construction of the upper bound function of the filtering error covariance. Then, we will continue to discuss the relationship between the energy level transmission probability and the trace of filtering error covariance.
			\end{remark}

    \subsection{Monotonicity of the activated probability}
        It should be pointed out that the trace of the filtering error covariance is an important performance index for designing the estimation gains, and the relationship between the activated probability and the performance index will be discussed below to evaluate the proposed $\hbar$+1-step filter.
		
		\begin{theorem}\label{th4}
            For the proposed $\hbar$+1-step unbiased filter \eqref{CW3b:1}-\eqref{CW3b:3}, the performance index $tr\{\Xi_u^\gamma(l,k)\}$ is non-increasing with increased $\varrho_{\hbar+1-\gamma}(l,k)$.
		\end{theorem}
		\begin{IEEEproof}
            Reviewing the activated  probability of energy storage shows that $\tilde{\varrho}_{\hbar}(l,k)=\varrho_{\hbar}(l,k)(1-\varrho_{\hbar}(l,k))$. It can be seen that
            \begin{small}
			\begin{align*}
                \bar{R}_{\hbar}(l,k)\triangleq&\varrho_{\hbar}^2(l,k)C_{\hbar}(l,k)\Xi_p^1(l,k)C_{\hbar}^\mathrm{T}(l,k)+\varrho_{\hbar}(l,k)\nonumber\\
				&\times(1-\varrho_{\hbar}(l,k))I\circ(C_{\hbar}(l,k)X(l,k)C_{\hbar}^\mathrm{T}(l,k))\nonumber\\
                &+\varrho_{\hbar}^2(l,k)\big(R_{\hbar}(l,k)+\sum_{\mu=1}^{r}\Delta_{h_\hbar,\mu}tr\{X(l,k)\Sigma_{h_{\hbar},\mu}\}\big)
			\end{align*}
            \end{small}
			In light of the property of trace derivation in \cite{Petersen12}, we know that
            \begin{small}
			\begin{align*}
                &\frac{dtr\big\{\Xi_u^1(l,k)\}}{d\{\varrho_{\hbar}(l,k)\big\}}=\frac{tr\big\{d\Xi_u^1(l,k)}{d\varrho_{\hbar}(l,k)\big\}}\nonumber\\
                &\quad=tr\big\{-2\varrho_{\hbar}(l,k)\Xi_p^1(l,k)C_\hbar^\mathrm{T}\bar{R}_\hbar^{-1}(l,k)C_\hbar(l,k)\Xi_p^1(l,k)\nonumber\\
                &\quad\ +\varrho_{\hbar}^2(l,k)\Xi_p^1(l,k)C_\hbar^\mathrm{T}\bar{R}_\hbar^{-1}(l,k)\big[2\varrho_{\hbar}C_\hbar(l,k)\Xi_p^1(l,k)\nonumber\\
                &\quad\ \times C_\hbar^\mathrm{T}(l,k)+(1-2\varrho_{\hbar}(l,k))I\circ(C_\hbar(l,k)X(l,k)C_\hbar^\mathrm{T}(l,k))\nonumber\\
                &\quad\ +2\varrho_{\hbar}(l,k)\big(R_{\hbar}(l,k)+\sum_{\mu=1}^{r}\Delta_{h_\hbar,\mu}tr\{X(l,k)\Sigma_{h_{\hbar},\mu}\}\big)\big]\nonumber\\
                &\quad\times\bar{R}^{-1}_\hbar(l,k)C_{\hbar}(l,k)\Xi_p^1(l,k)\big\}\nonumber\\
                &\quad=-\varrho_{\hbar}^{-1}(l,k)tr\big\{K_1(l,k)\{2\bar{R}_\hbar(i,j)-2\varrho_{\hbar}^2(l,k)C_{\hbar}(l,k)\nonumber\\
                &\quad\ \times\Xi_p^1(l,k)C_{\hbar}^\mathrm{T}(l,k)-\varrho_{\hbar}(l,k)(1-2\varrho_{\hbar}(l,k))I\nonumber\\
				&\quad\ \circ(C_{\hbar}(l,k)X(l,k)C_\hbar^\mathrm{T})\}K_1^\mathrm{T}(l,k)\big\}\nonumber\\
                &\quad=-\varrho_{\hbar}^{-1}(l,k)tr\big\{K_1(l,k)\{\varrho_{\hbar}(l,k)I\circ(C_\hbar(l,k)X(l,k)C_{\hbar}^\mathrm{T}(l,k))\nonumber\\
				&\quad\  +2R_{\hbar}(l,k)+\sum_{\mu=1}^{r}\Delta_{h_\hbar,\mu}tr\{X(l,k)\Sigma_{h_{\hbar},\mu}\}\big\}K_1^\mathrm{T}(l,k)\}\nonumber\\
				&\quad\leq0
			\end{align*}
            \end{small}
            This completes the proof.
			\end{IEEEproof}
			\begin{remark}
                It can be seen that the greater the activated probability, it indicates that the energy storage of the sensor is sufficient, and the authorized transmission data between the sensor and the remote filter is frequent. As the number of available activated observations increases, the better the filtering performance is expected to be, the smaller the trace of the filtering error covariance is reflected. This is consistent with the theoretical analysis of Theorem~\ref{th4}.
			\end{remark}

\section{Conclusion}\label{sec:6}
    This paper mainly studies the state estimation problem for shift-varying 2-D systems with asynchronous multi-channel delays and energy harvesting constraints. The energy harvesting constraints considered are described by whether the energy storage of the wireless sensor the authorized transmission of the sensor to the remote filter. To effectively utilize the delayed observations, an observation reconstruction approach have been proposed. Based on the reconstructed observations and activation probability, an $\hbar$+1-step recursive filter have been constructed. Then, the required estimation gains has been recursively calculated to ensure that the corresponding filter error covariance have been minimized. Then, the boundedness analysis of the estimation error covariance has been given. And the monotonicity analysis of the activated probability on estimation performance has been given. Finally, the effectiveness of the designed estimator is verified through a numerical simulation example. Future research topics include: 1) distributed state estimation problem with asynchronous multi-channel delays and energy harvesting constraints and 2) asynchronous random delays based state estimation problem.

\end{document}